\documentclass[fleqn,usenatbib]{mnras}

\usepackage{newtxtext,newtxmath}
\usepackage[T1]{fontenc}
\usepackage{ae,aecompl}

\usepackage{graphicx}	% Including figure files
\usepackage{amsmath}	% Advanced maths commands
\usepackage[dvipsnames,usenames]{color} 
\usepackage{longtable}
\usepackage{float} 
\usepackage{natbib}
\usepackage{adjustbox}
\usepackage[normalem]{ulem}

\bibpunct{(}{)}{;}{a}{}{,}

\title[PLATO Hare and Hounds for MS stars]{{PLATO Hare-and-Hounds exercise: {Asteroseismic model fitting of main-sequence solar-like pulsators}}}
\author[M.S. Cunha et al.]{M.~S.~Cunha,$^{1}$\thanks{E-mail: mcunha@astro.up.pt}
I.~W.~Roxburgh,$^{2,3}$ V. Aguirre B\o rsen-Koch,$^{4}$ W.~H.~Ball,$^{3,4}$ S.~Basu,$^{5}$ \newauthor W.~J.~Chaplin,$^{3,4}$  M.-J.~Goupil,$^{6}$  B.~Nsamba,$^{7,1,8}$ J.~Ong,$^{5}$ D.~R.~Reese,$^{6}$ K.~Verma,$^{4}$\newauthor K.~Belkacem,$^{6}$ T.~Campante,$^{1,9}$ J.~Christensen-Dalsgaard,$^{4}$ M.~T.~Clara,$^{1,9}$ \newauthor S.~Deheuvels,$^{10}$ M.~J.~P.~F.~G.~Monteiro,$^{1,9}$  A.~Noll,$^{10}$ R.~M.~Ouazzani,$^{6}$ J.~L.~Rørsted,$^{4}$ \newauthor A.~Stokholm,$^{4,11}$ M.~L.~Winther$^{4}$
\\
$^{1}$ Instituto de Astrof\'\i sica e Ci\^encias do Espa\c co, Universidade do
Porto, CAUP, Rua das Estrelas, PT4150-762 Porto, Portugal\\
$^{2}$ Astronomy Unit, Queen Mary University of London, Mile End Road, London E1 4NS, UK\\
$^{3}$ School of Physics and Astronomy, University of Birmingham, Edgbaston, Birmingham B15 2TT, UK\\
$^{4}$ Stellar Astrophysics Centre, Department of Physics and Astronomy, Aarhus University, Ny Munkegade 120, DK-8000 Aarhus C, Denmark\\
$^{5}$ Department of Astronomy, Yale University, PO Box 208101, New Haven, CT 06520-8101, USA\\
$^{6}$ LESIA, Observatoire de Paris, Université PSL, CNRS, Sorbonne Université, Université de Paris, 5 Place Jules Janssen, 92195, Meudon, France \\
$^{7}$ Max-Planck-Institut f\"{u}r Astrophysik, Karl-Schwarzschild-Str. 1, D-85748 Garching, Germany\\
$^{8}$ Kyambogo University, P.O. Box 1, Kyambogo, Kampala - Uganda\\
$^{9}$ Departamento de F\'isica e Astronomia, Faculdade de Ci\^encias da Universidade do Porto, Rua do Campo Alegre, s/n, PT-4169-007 Porto, Portugal\\
$^{10}$ IRAP, Université de Toulouse, CNRS, CNES, UPS, Toulouse, France\\
$^{11}$ Dipartimento di Fisica e Astronomia, Università degli Studi di Bologna, Via Gobetti 93/2, I-40129 Bologna, Italy\\
}

\date{Accepted XXX. Received YYY; in original form ZZZ}

\pubyear{2015}

\begin{document}
\label{firstpage}
\pagerange{\pageref{firstpage}--\pageref{lastpage}}
\maketitle

\begin{abstract}
{Asteroseismology is a powerful tool to infer fundamental stellar properties. The use of these asteroseismic-inferred properties in a growing number of astrophysical contexts makes it vital to understand their accuracy. Consequently, we performed a hare-and-hounds exercise where the hares simulated data for 6 artificial main-sequence stars and the hounds inferred their properties based on different inference procedures. To mimic a pipeline such as that {planned} for the PLATO mission, all hounds used the same model grid. Some stars were simulated using the physics adopted in the grid, others a different one. The maximum relative differences found (in absolute value) between the inferred and true values of the mass, radius, and age were 4.32~per~cent, 1.33~per~cent, and 11.25~per~cent, respectively. The largest systematic differences in radius and age were found for a star simulated assuming gravitational settling, not accounted for in the model grid, with biases of -0.88~per~cent (radius) and 8.66~per~cent (age). For the mass, the most significant bias (-3.16~per~cent) was found for a star with {a helium enrichment} ratio outside the grid range. {Moreover, a $\sim$7~per~cent dispersion in age was found when adopting different prescriptions for the surface corrections or shifting the classical observations by $\pm 1\sigma$. The choice of the relative weight given to the classical and seismic constraints also impacted significantly the accuracy and precision of the results. Interestingly, only a few frequencies were required to achieve accurate results on the mass and radius. For the age the same was true when at least one $l=2$ mode was considered.}}
\end{abstract}

\begin{keywords}
{asteroseismology -- stars: fundamental parameters -- stars: evolution -- stars: oscillations -- methods: statistical}
\end{keywords}

%%%%%%%%%%%%%%%%%%%%%%%%%%%%%%%%%%%%%%%%%%%%%%%%%%

%%%%%%%%%%%%%%%%% BODY OF PAPER %%%%%%%%%%%%%%%%%%

\section{Introduction}
\label{int}
%\red{[Something about PLATO; limitations concerning the modelling of the physics and dynamics of stars and their evolution; diversity of forward modelling techniques (use of different observables and weighting, different ways of exploring parameters space - grid-based, model-on-the-fly -  surface independent/dependent techniques); something on systematic uncertainties. Mention that there is a very limited number of benchmark stars for asteroseismology and argue for the need for controlled experiments (HH). End with the goals set for this exercise: test different FM techniques including the different aspects mentioned before, for a fixed grid of models (as will be the case of PLATO); verify the impact on the inferred parameters from changing the uncertainties in the classical observations. Mention that tests include also stars that are "outside" the grid, either because of the physics adopted for the grid or because of the parameter space covered; ]}

{Stellar characterisation is a matter of fundamental importance in the general astrophysical context. Exoplanet research~\citep[e.g.][]{winn15,santos18} and {Galactic} archaeology~\citep[e.g.][]{miglio17} are examples of areas where studies often rely on the knowledge of fundamental stellar properties, such as the stellar mass, radius, and age. The advent of space-based asteroseismology has greatly enhanced the precision with which these stellar properties can be inferred \citep{chaplin14,legacy2017}, leading to strong and long-lasting synergies between asteroseismology and these other fields of research. An example of such synergy is provided by the ESA mission PLAnetary Transits and Oscillations of stars (PLATO)~\citep{rauer14}, where the hunt for terrestrial planets is planned to go hand-in-hand with the characterisation of their host stars through asteroseismology. In this context, it is fundamental to understand to what precision and accuracy stellar properties may be derived from space-based asteroseismic data such as that planned to be acquired by PLATO. }

{Earlier works based on data collected by the Kepler satellite~\citep{gillilandetal10} have been particularly informative concerning the precision of asteroseismic-inferred stellar properties. \cite{chaplin14} showed that access to just two seismic global constraints, namely, the frequency of maximum oscillation power $\nu_{\rm max}$ and the large frequency separation $\Delta\nu$, enables the inference of stellar masses, radii, and ages with typical uncertainties of $\sim$5.4 per cent, $\sim$2.2 per cent, and $\sim$25 per cent, respectively, when spectroscopic constraints are simultaneously available. These uncertainties are further reduced to averages of  $\sim$4 per cent in mass, $\sim$2 per cent in radius, and $\sim$10 per cent in age, when a significant number of individual mode frequencies are detected, as shown by \cite{legacy2017} in a study of the 66 stars in the Kepler Legacy sample.  {Importantly, in both studies the uncertainties quoted are not the statistical errors from a single pipeline, but consider the results from different evolutionary codes combined with a variety of model physics, and different analysis methods.}}

{While the Kepler legacy is extremely valuable in the context of the preparation for the PLATO mission, the results presented in the works mentioned above do not inform us on the accuracy of the asteroseismic inferences. Consistency checks against the results from independent methods are possible in some cases~\citep{bruntt10,huber12,sahlholdt18}. However, to truly test the accuracy of the asteroseismic results one would need access to independently-derived stellar properties whose statistical and systematic errors are significantly smaller than the uncertainties on the asteroseismic inferences. That may be possible for the mass and radius, from the study of eclipsing binaries \citep[e.g.][]{torres10,serenelli21} and for the mass alone, from the study of some double-lined spectroscopic binaries \citep[e.g.][]{halbwachs20}. Unfortunately, with very few exceptions, at the present date such accurate measurements are not available for stars having, simultaneously,  asteroseismic data. Asteroseismic observation of a number of such benchmark stars following the launch of PLATO should enable future tests to the accuracy of the asteroseismic inferences.  }

{An alternative way to access the accuracy of the asteroseismic inference procedures is to resort to simulated data. Any tests based on simulated data are limited by one's ability to produce realistic representations of the real data sets. Therefore, they cannot evaluate the impact of physical processes not included in the models used to simulate the data that may be at play in stars. Nevertheless, these tests are useful to understand the biases that are introduced in the inferred stellar properties by known sources of systematic errors, {which} can be accounted for in the simulations. Exercises of this type have been performed earlier both based on simulated data sets including only global seismic observations~\citep{stello09} and simulated data sets including individual-mode frequencies~\citep{reese16}. Nevertheless, in both cases the underlying stellar models and associated models' physics varied according to the modeller's choice, hindering a direct comparison of the different inference procedures.} 

{In this work we use simulated data to establish the {accuracy limit} with which stellar properties may be derived from given sets of asteroseismic data. Our goal is to compare the performances of different grid-based inference methods used by the asteroseismic community. Specifically, we perform a hare-and-hounds exercise, where the hares produce simulated data for a set of targets and the hounds try to recover the true properties of these targets.  All hounds were asked to use the same grid of stellar models and frequencies, such as to mimic the future PLATO pipeline. Consequently, the differences in the inferences made by different hounds result solely from the differences in the methods employed. Nevertheless, some of the targets were simulated using a physics setup differing from the one used to build the grid of models, or adopting parameter values outside the grid parameter space. Therefore, in those cases, the differences between the inferred values and the true values reflect also the biases that are introduced in the grid-based inference problem when fixing the physics of the models in a grid.}

{The {remainder} of this paper is organised as follows: in Section~\ref{HH} we introduce the hare-and-hounds exercise, specifying the characteristics of the grid of models adopted for the inferences, the properties of the simulated stars and the simulation procedure.  Section~\ref{met} highlights the main differences between the grid-based inference methods considered in the exercise. Section~\ref{res} discusses the results from the exercise,  comparing the inferences made based on different procedures. Sections~\ref{sec:surf} to \ref{errors} then assess the impact on the results from considering different prescriptions for the surface corrections, changing the relative weight given to the classical and seismic observations, degrading the quality of the seismic data, and shifting the uncertainties in the classical observations. Finally, in Section~\ref{sec:conclusions} we summarise our conclusions. }

\section{Setting the experiment}
\label{HH}
\subsection{Grid of models}
\label{grid}
We used the Modules for Experiments in Stellar Astrophysics \citep[MESA version 10108;][]{paxt11,paxt13,paxt15,paxt18} to compute the stellar model grid.
The MESA code provides several options for various input physics. We used it with Opacity Project (OP) high-temperature opacities \citep{badn05,seat05} 
supplemented with low-temperature opacities of \citet{ferg05}. The metallicity mixture from \citet{gs98} was used. We used {the} OPAL equation of state 
\citep{roge02}. The reaction rates were from NACRE \citep{angu99} for all reactions except $^{14}{\rm N}(p, \gamma)^{15}{\rm O}$ and 
$^{12}{\rm C}(\alpha, \gamma)^{16}{\rm O}$, for which updated reaction rates from \citet{imbr05} and \citet{kunz02} were used, respectively. For overshoot, 
we used the prescription of \citet{herw00}. The Eddington $T-\tau$ relation \citep{eddi26} was used for atmospheric boundary conditions. {The initial helium mass fraction, $Y_{\rm ini}$, was derived from the initial metal mass fraction, $Z_{\rm ini}$, through a helium-to-heavy metal enrichment law, 
\begin{equation}
    Y_{\rm ini}=\frac{dY}{dZ}Z_{\rm ini}+Y_0,
\end{equation}
with a {Big Bang} nucleosynthesis value for the helium mass fraction of $Y_0=0.248$.} {The formalism for convection was used from \citet{cox68}.}  The model 
oscillation frequencies, $\nu_{nl}$,  where  $n$ is the radial order and $l$ the degree, were calculated using the Aarhus adiabatic oscillation package \citep[ADIPLS;][]{jcd08a} with isothermal atmosphere boundary 
condition. 

We generated a hybrid stellar model grid with a total of 9000 evolutionary tracks {containing about 3.5 million models}; the mass and initial metallicity were sampled uniformly in predefined ranges ($M \in [0.8, 1.5]$ M$_\odot$ and 
$[{\rm Fe}/{\rm H}]_i \in [-0.5, 0.5]$ dex) using a quasi-random number generator \citep{sobo67}, whereas mixing-length, overshoot and helium-to-metal 
enrichment ratio were sampled uniformly from predefined sets of values ($\alpha_{\rm mlt} \in \{1.6, 1.7, 1.8, 1.9, 2.0\}$, 
$f_{\rm ov} \in \{0.0, 0.015, 0.030\}$ and $dY/dZ \in \{1, 2, 3\}$). {The model profiles have about 2000 mesh points.}

%\subsection{Hares}
\subsection{Simulated stars}
\label{hares}

\begin{figure}
	\includegraphics[width=1\columnwidth]{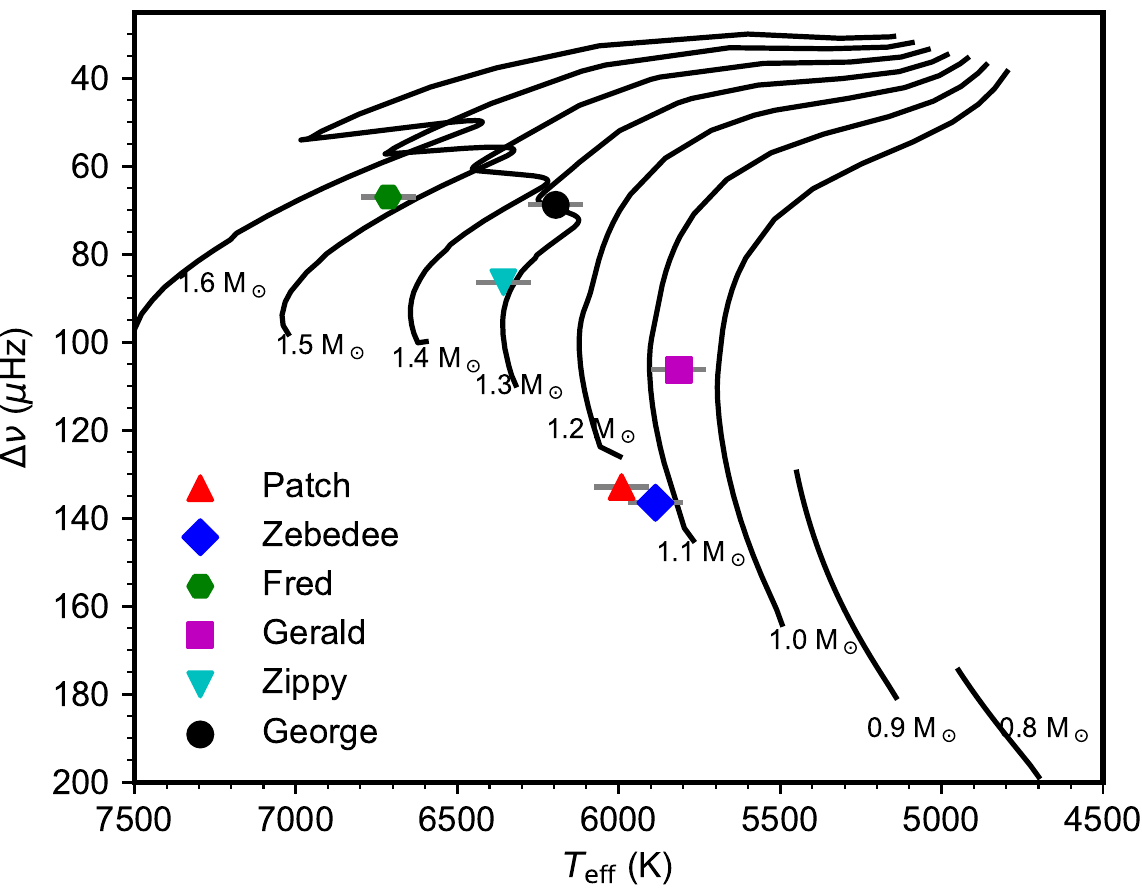}
    \caption{{Location in the asteroseismic HR~diagram of the six targets produced by the hares. {The uncertainties in $T_{\rm eff}$ are indicated by grey horizontal bars while the uncertainties in $\Delta\nu$ are smaller than the symbols.} Solar metallicity evolutionary tracks (black lines) with masses in the range 0.8 - 1.6 M$_\odot$ constructed using a mixing length parameter ($\alpha_{\rm{mlt}}$) of 1.8 and without element diffusion,
    are also shown for guidance.} 
    }
    \label{fig:HR}
\end{figure}

{The hares produced data for six simulated stars - hereafter, the targets - named Patch, Zebedee, Fred, Gerald, Zippy, and George. Their location in the HR~diagram is shown in Fig.~\ref{fig:HR}.} Their main properties are listed in Table~\ref{tab:hares} and their simulated classical and global seismic constraints are shown in Table~\ref{tab:classical}. The simulated individual mode frequencies are listed in Tables~\ref{tab:freq_hares} and \ref{tab:freq_hares_2}, in Appendix~\ref{apA}.

 Patch, Zebedee, and Fred were generated with the default physics used to construct the grid (Section~\ref{grid}), but parameters 
 %for mixing length, overshooting, and the enrichment ratio 
 were allowed to differ from the values in the grid. For the other three {targets}, the adopted model physics has been modified as follows. For George, we used an atmospheric $T-\tau$ relation fit to the empirical solar atmosphere model C by \cite{vernazza81} implemented in MESA as the solar${\_}$Hopf${\_}$grey option (see Sec. A.5 of \cite{paxt13}).  For Zippy, we included convective overshoot with a step-like diffusion profile at the convective core boundary only, rather than the exponentially-decaying profile at all boundaries used in the other models.  The diffusion coefficient of convective mixing was extended from 0.001$H_{\rm p}$ below the convective boundary to 0.2$H_{\rm p}$ above{, where $H_{\rm p}$ is the pressure scale height}.  For Gerald, we included the effects of gravitational settling implemented using the method by \cite{thoul94}. Finally, for two of the {targets}, one of the parameters was beyond the grid limits. In particular, for Fred, an enrichment ratio of $dY/dZ=0.77$ was adopted and for George the overshoot was taken to be $f_{\rm ov}=0.0939$. 
 
%\red{[Bill: Added below...}
 
 To produce the artificial observations -- i.e., individual mode
frequencies, global asteroseismic parameters, and their uncertainties,
for each star -- we followed the approach and recipe of
\cite{reese16}. Full details of the procedures may be found in that
paper, but to summarize: The fundamental properties of each artificial
star were used as input to scaling relations, which calculated the
expected underlying parameters of the oscillation spectrum and the
intrinsic background arising from granulation. For the base exercise,
all artificial stars were assumed to be observed {continuously for a period of 2yr} at an apparent visual
magnitude of $V=9$, which, coupled to a model for the PLATO noise
performance defined the expected {noise} level for each star. With
the appearance of the underlying (so-called limit) spectrum defined,
we used analytical relations to calculate the probability of detection
for each mode, and the expected precision in their frequencies.

Frequencies of those modes flagged as detectable were perturbed by
adding a random Gaussian {perturbation} of standard deviation equal to the
expected frequency precision, and passed to the list of simulated
observed outputs. This list was augmented by observational estimates
of the global asteroseismic parameters $\nu_{\rm max}$ and $\Delta\nu$, both
computed using scaling relations, with the central values perturbed
based on assumed measured precisions of 5\,\% and 2\,\%, respectively.

The non-seismic observations -- luminosity $L$, effective temperature
$T_{\rm eff}$, and metallicity [Fe/H] -- were created in a similar
manner, by perturbing the true values assuming measured precisions of
3\,\%, 85\,K and 0.09\,dex, respectively.
 
% \red{[Warrick: can you please check the text below and see if it needs changing in light of Joergen's comment (his email from 06/07)?]}
{Finally, to mimic the systematic differences known to exist between model and observed frequencies as a result of the deficient modelling of the surface layers of stars,  a surface effect was added to the artificial frequency data using the one-term (or ``cubic" correction) by \cite{ball14} (hereafter BG-1term).  We started with the coefficient found by fitting Model S \citep{jcd96} to low-degree mode frequencies from BiSON  \citep{broomhall09,davies14} and, for each simulated star, multiplied the coefficient by a random number drawn uniformly between 0.98 and 1.02.}

\begin{table*}
	\centering
	\caption{Properties of the {targets}.}
	\label{tab:hares}
	\resizebox{2\columnwidth}{!}{%
	\begin{tabular}{lccccccccll} % four columns, alignment for each
		\hline
{\bf {Targets}} & ID &Mass (M$_{\odot}$) & Radius (R$_{\odot}$) & Age (Gyr) & $Y_{\rm ini}$ & $Z_{\rm ini}$ & $\alpha_{\rm mlt}$ & $f_{\rm ov}$ & Physics & Notes \\
		\hline
		Patch & Pa &0.8644 & 0.9557 & 9.898 & 0.25906 & 0.00784 & 1.931 & 0.0115 & Default &\\
		Zebedee & Ze & 1.0165 & 0.9646 & 3.085 & 0.26786 & 0.01734 & 1.872 & 0.0223 & Default &\\
		Fred & Fr &1.4318 & 1.7225 & 1.839 & 0.26055 & 0.01638 & 1.688 & 0.0066 & Default & $dY/dZ$$^*$\\
		Gerald & Ge & 1.0242 & 1.2053 & 8.039 & 0.27566 & 0.02111 & 1.967 & 0.0274 & Gravitational settling \\
		Zippy & Zi & 1.1278 & 1.3965 & 4.223 & 0.27784 & 0.01245 & 1.880 & -- & Step overshooting$^+$ \\
		George & Go & 1.3430 & 1.7069 & 3.757 & 0.28049 & 0.03001 & 1.770 & 0.0939 & VAL C atmosphere  & $f_{\rm ov} ^*$ \\
		\hline
		\multicolumn{11}{l}{$^*$ The value is outside the grid parameter space.} \\
		\multicolumn{11}{l}{$^+$  See Section~\ref{hares} for details.} \\
	\end{tabular}%
	}
\end{table*}

\begin{table*}
	\centering
	\caption{{Classical and global seismic parameters of the {targets}. The  model luminosity, effective temperature, and surface iron abundance, as well as the model $\nu_{\rm max}$ and $\Delta\nu$ determined through the scaling relations, are marked by the superscript "true". Each of these quantities is followed by the simulated value and 1\,$\sigma$ error provided to the hounds (see text for details). }}
	\label{tab:classical}
	\resizebox{2.0\columnwidth}{!}{%
	\begin{tabular}{lcccccccccc} % four columns, alignment for each
		\hline
{\bf {Targets}} 	&	 $L/L_\odot^{\rm true}$	&	 $L/L_\odot$ 	&	$T_{\rm eff}^{\rm true}$ (K) 	&	 $T_{\rm eff}$ (K) 	&	 [Fe/H]$^{\rm true}$ 	&	 [Fe/H] 	&	 $\nu_{\rm max}^{\rm true}$ ($\mu$Hz)  	&	 $\nu_{\rm max}$ ($\mu$Hz) 	&	 $\Delta\nu^{\rm true}$ ($\mu$Hz) 	&	  $\Delta\nu$ ($\mu$Hz) 	\\
\hline																					
Patch 	&	1.0737	&	 1.03 $\pm$ 0.03 	&	6014.4260	&	 5991 $\pm$ 85 	&	-0.3329	&	 -0.28 $\pm$ 0.09 	&	2865	&	 2906 $\pm$ 143 	&	134.4	&	 132.9 $\pm$ 2.7 	\\
Zebedee 	&	0.9982	&	 0.98 $\pm$ 0.03 	&	5878.4143	&	 5886 $\pm$ 85 	&	0.0238	&	 0.10 $\pm$ 0.09 	&	3345	&	 3254 $\pm$ 167 	&	143.8	&	 136.5 $\pm$ 2.8 	\\
Fred 	&	5.3753	&	 5.42 $\pm$ 0.16 	&	6701.0619	&	 6714 $\pm$ 85 	&	-0.006	&	 -0.04 $\pm$ 0.09 	&	1384	&	 1393 $\pm$ 69 	&	71.5	&	 67.0 $\pm$ 1.4 	\\
Gerald 	&	1.5481	&	 1.50 $\pm$ 0.05 	&	5868.5382	&	 5814 $\pm$ 85 	&	0.0375	&	 0.03 $\pm$ 0.09 	&	2160	&	 2207 $\pm$ 108 	&	103.3	&	 106.3 $\pm$ 2.1 	\\
Zippy 	&	2.8445	&	 2.85 $\pm$ 0.09 	&	6347.5330	&	 6357 $\pm$ 85 	&	-0.1172	&	 -0.17 $\pm$ 0.09 	&	1704	&	 1660 $\pm$ 85 	&	86.9	&	 86.4 $\pm$ 1.7 	\\
George 	&	3.8169	&	 3.67 $\pm$ 0.11 	&	6179.4857	&	 6195 $\pm$ 85 	&	0.2779	&	 0.35 $\pm$ 0.09 	&	1377	&	 1284 $\pm$ 68 	&	70.2	&	 68.8 $\pm$ 1.4 	\\
\hline																					
	\end{tabular}%
	}
\end{table*}

\section{Methods}
\label{met}

The {targets}' properties were inferred by five modellers, hereafter, the hounds, through a series of grid-based inference methods. All hounds used the same grid of models and frequencies (described in Section~\ref{grid}). The goal was to understand how the differences in the optimisation methods employed by the hounds to explore the grid impact on the results. {The hounds produced probability distributions for the stellar properties reporting, in each case, the mean of the distribution, a 1$\sigma$ uncertainty on the mean and the values of the 16th, 50th and 84th percentiles.}  Some hounds submitted different sets of results that were either inferred with different methods or with the same method but applying different weights to the observations or different prescriptions for the surface corrections. In those cases, one method and one associated set of results was elected for the comparison discussed in Section~\ref{sec:comparison}, prior to the true values of the {targets}' properties being revealed. The elected methods were chosen such as to guarantee that the approaches showcased were as diverse as possible. The list of hounds is presented in Table~\ref{tab:hounds} and the detailed description of the methods is presented in Appendix~\ref{apB}. 

%along with the main characteristics of the elected method used for the inferences reported in Section~\ref{sec:comparison}. 
\subsection{Key differences}
\label{sec:differeces}

\begin{table*}
	\centering
	\caption{Inference methods employed by the different hounds. The first five rows concern the elected methods compared in Fig.~\ref{fig:comparison}. The additional methods listed (IR$_\nu$ and VA$_{\rm int}$) are variants used in specific tests only (see text for details). A detailed description of the methods is provided in Appendix~\ref{apB}.
	%\red{[To solve: Need to check which name hounds want to adopt. We could keep the initials and add a column for those that have a named code (e.g., BASTA, AIMS, etc).]}
	}
	\label{tab:hounds}
	\resizebox{2.0\columnwidth}{!}{%
	\begin{tabular}{llllll} % four columns, alignment for each
		\hline
{\bf Hounds ID} & Colour & Surface/correction$^*$ & Observations & weights$^+$ & Interpolation/Sampling  \\
		\hline
		SB & Black &  dependent/BG-2term & $T_{\rm eff}$,$L$,[Fe/H], $\nu_{nl}$ & 3:1 $\&$ 2 lowest $\nu_{nl}$ (${\rm all}~l$)  & no/Grid\\
%		BN & Red &   dependent & & & yes \\
		JO & Green &   dependent/BG-2term  & $T_{\rm eff}$,$L$,[Fe/H],$\nu_{\rm max}$,$\nu_{nl}$,$\epsilon_l(\nu_{nl})$$^\dagger$ & {5:2} $\&$ 5 lowest $\nu_{n0}$ [3:3] & no/Grid \\
		DR & Blue &  dependent/Various & $T_{\rm eff}$,$L$,[Fe/H],$\nu_{nl}$  & 3:3 [3:1;3:N] & yes/MCMC \\
		IR$_{\epsilon}$ & Brown &   independent & $T_{\rm eff}$,$L$,[Fe/H],$\epsilon_l(\nu_{nl})$&  3:3 [3:1;3:N] $\&$ lowest $\nu_{n0}$ & no/Grid \\
		VA & Magenta &  dependent/BG-2term & $T_{\rm eff}$,$L$,[Fe/H], $\nu_{nl}$ &  3:1 [3:3;3:N] & no/Grid\\
		\hline
		IR$_\nu$ & Orange & dependent/Various &  $T_{\rm eff}$,$L$,[Fe/H], $\nu_{nl}$  &3:3 [3:1;3:N] & no/Grid\\
		VA$_{\rm int}$ & -- &  dependent/BG-2term & $T_{\rm eff}$,$L$,[Fe/H], $\nu_{nl}$ &  3:1 & yes/{20}xGrid\\
		\hline
	    \multicolumn{6}{l}{$^*$ Whenever various surface corrections are considered, the elected case (Fig.~\ref{fig:comparison}) adopted the \cite{ball14} two-term correction (BG-2term).}\\ 
	    \multicolumn{6}{l}{$^+$ Whenever several weights are listed, the one adopted for the elected case (Fig.~\ref{fig:comparison}) is shown outside the squared brackets.} \\
	    \multicolumn{6}{l}{$^\dagger$ For the results shown in Fig.~\ref{fig:pdfs}, only a subset of these observations  was considered, namely: $T_{\rm eff}$, $L$, [Fe/H] and $\nu_{nl}$.}\\
	    % JO: looks good to me
	\end{tabular}%
	}
\end{table*}

{The inference methods discussed in this work differ in a few key aspects. One of these concerns the way the parameter space is sampled. In most cases, the sampling is limited to the grid points. In one case (variant VA$_{\rm int}$ in Table~\ref{tab:hounds}) interpolation is carried out prior to the fitting, such that the number of evolutionary tracks is increased by a factor of twenty with respect to the original grid, and the frequency resolution along any given track is increased to guarantee a maximum of 1$\mu$Hz variation of the $l=0$ mode of lowest radial order observed  between consecutive models. The interpolation is performed in a region of the grid selected according to the observed values of effective temperature, metallicity, and large frequency separation. 
%\red{[Victor: is this it? Is interpolation made on the whole grid, or just in a selected region? is there any iteration? Please adjust the text as you see fit.]}. 
In all these methods, the seismic and classical constraints are fitted to the corresponding data counterparts at each grid point (or at a subset of those), being it the original grid, or the grid that follows from the interpolation. In contrast, in one case (DR) the sampling is based on a Markov Chain Monte Carlo (MCMC) approach {\citep[e.g.][]{metropolis1953, hastings1970}}. Here, the model observables also need to be computed between grid points, which is again achieved  with recourse to interpolation.}

{Another aspect in which the inference methods may differ concerns the way the stellar properties and their uncertainties are computed. In most cases they are derived directly from the mass, radius, and age probability distributions inferred from the fits. However, in one method (JO), Monte Carlo (MC) simulations are performed by varying the non-seismic and global asteroseismic observations within their errors. In each simulation, the means of the probability distributions are collected to build distributions for the mean values. The reported
values and uncertainties for the mass, radius, and age are then derived from the probability distributions of the posterior means.}

{In addition to the above, depending on the seismic quantities considered in the fits, the methods may be {considered} surface dependent or independent, {in the sense that they may either include or not include a parametrized surface correction to the model frequencies}. In most cases, the individual observed frequencies were fitted to the model counterparts. In order to proceed this way, the model frequencies were first corrected for surface effects
%, a systematic and frequency-dependent frequency shift resulting from the incorrect modelling of structural and dynamical aspects of the near-surface layers of a star, as well as of the oscillations  
\citep{kjeldsen08,ball14,sonoi15}.
While having the advantage of setting significant constraints on global properties such as the stellar radius and mean density, inferences based on fitting individual frequencies may be subject to biases associated with {a possibly improper treatment} of surface effects. An alternative provided by one of the methods (IR$_\epsilon$) is to apply a {surface-independent} approach, by which the seismic data is first combined in such a way as to produce a new set of data (in this case, the phases $\epsilon_{l}$; see Appendix~\ref{apB} for {a}  definition) that enables the search for models with an interior structure similar to that of the star, without having to parametrize the effect of the outer layers on the seismic data \citep{roxburgh03,roxburgh15,roxburgh16}. As a consequence of their limited {sensitivity} to the outer layers, surface-independent methods have little constraining power on the stellar radius and mean density. To overcome that, the frequency of the radial mode of lowest radial order is also fitted. As the surface correction is smallest at low frequencies, the expectation is that fitting this mode without employing a surface correction will provide enough additional information to the otherwise surface-independent method to constrain the stellar radius and density, without biasing the results.}

{For any given method, the hounds considered a set of observations to fit, including global and individual seismic constraints (individual frequencies and/or individual phases derived from those frequencies). In most cases, the global constraints consisted of $L$, $T_{\rm eff}$ and [Fe/H]. In one case (JO), $\nu_{\rm max}$ and $\epsilon_c$ (the radial mode phase offset at $\nu_{\rm max}$, in the sense of \cite{ong19}) were also added to the global constraints. For the chosen set of observations, the hounds then considered either one or several options for the relative weight given to the global and individual seismic constraints. For a case of a fit to three global constraints and N individual frequencies, a 3:N weight means that each of the observations is given the same weight, while a 3:3 weight means that the three global constraints together are given the same weight as the N individual frequencies together, and a 3:1 weight means that all N frequencies together are given the same weight as one global constraint. Whenever several options were considered for the weight, the one chosen for the method elected for comparison in Section~\ref{sec:comparison} is listed outside the square brackets in Table~\ref{tab:hounds}. }

\subsection{Impact on the probability distributions}
\label{sec:pdfs}

\begin{figure*}
	\includegraphics[width=1.9\columnwidth]{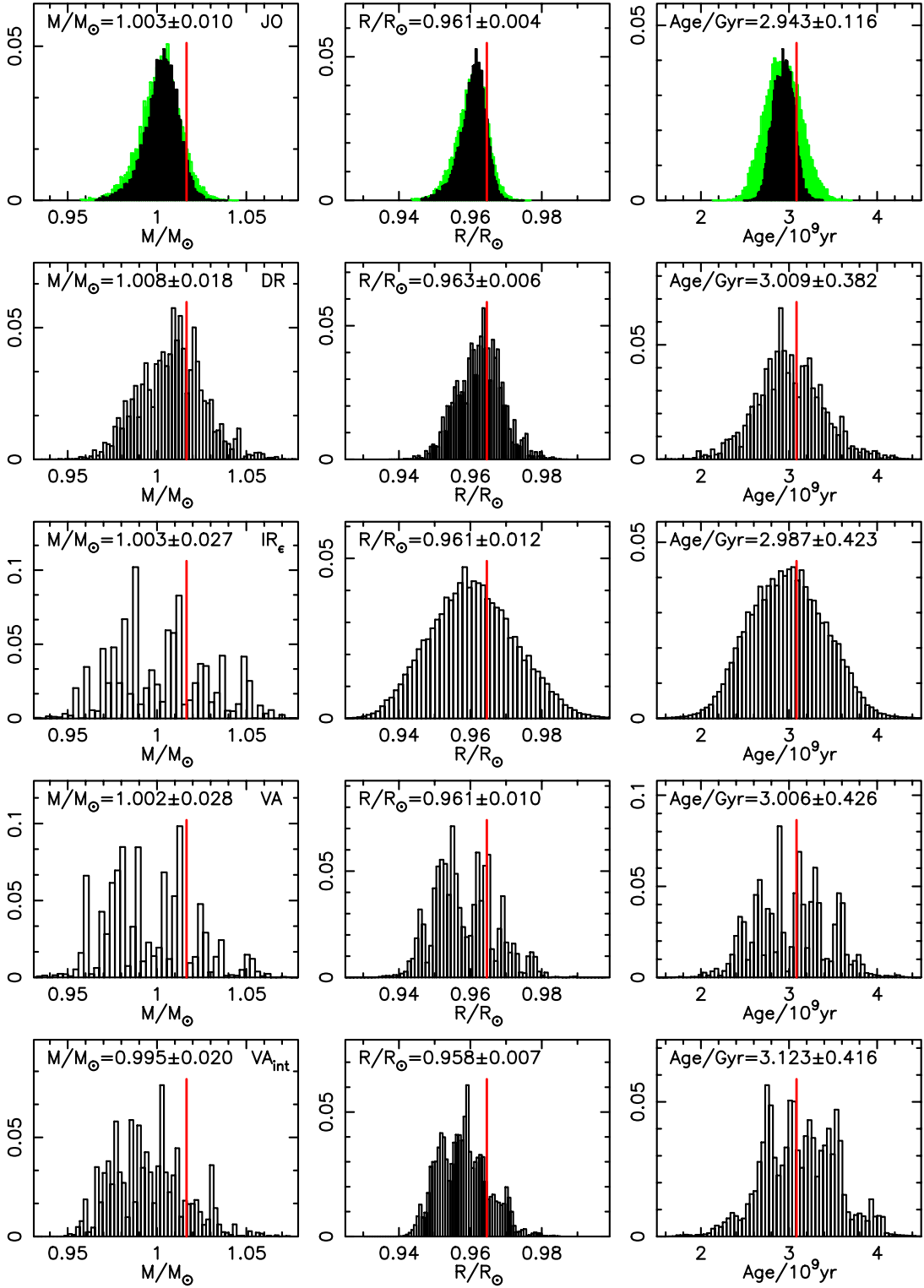}
   \caption{
   %\red{[Here the values are from the medians. Do we want to use that and take the opportunity to say the mean and median are not too different (as the distributions as well behaved)? Or should we use the mean, as in the table and figure 1, for consistency? If we chose the median, then we need to write the errors as the 16th and 84th percentiles and not half the difference between them].} 
   {Probability mass functions for the mass (left), radius (middle) and age (right) of Zebedee. From top to bottom, the rows show the results for four hounds, respectively: JO, DR, IR$_\epsilon$, VA. The bottom row shows the results of the hound VA when interpolation on the grid is considered (see text for details). {The green highlight in the top panels shows the distributions by JO when the frequencies are also perturbed in the MC simulations (see text for details).} All hounds applied a 3:3 weight and, where applicable, the BG-2term correction. The red vertical lines mark the true values of the parameters. The results shown inside the panels correspond to the mean and 1\,$\sigma$ uncertainties of the inferred properties (in the top panels the values are for the results in black{; for the results in green we found $M/$M$_\odot$=1.002 $\pm$ 0.011, $R/$R$_\odot$=0.960 $\pm$ 0.004 and Age/Gyr=2.934 $\pm$ 0.201}). We note that the probability mass function is defined for discrete variables. Here, that is achieved by binning each property in 76 bins (see text for details). Thus, it is the sum of the probability mass function values over all bins (rather than the area under the curve) that is equal to one.  }}
    \label{fig:pdfs}
\end{figure*}

{The key differences discussed above impact the probability distributions inferred from the fits. This is illustrated in Fig.~\ref{fig:pdfs} where the probability distributions inferred for the properties of Zebedee are shown for five different methods (JO, DR, IR$_\epsilon$, {VA,} and VA$_{\rm int}$). Here we chose to show  probability mass functions, which are defined for discrete variables. These were computed from the probability distributions for each property by considering an interval of $\pm 4\sigma$ centred on the mean value, binning in 76 equal-size bins, and normalising, such that the probability for each bin and property can be directly read from the corresponding y axis in Fig.~\ref{fig:pdfs}. 

To assure that the differences in the inferences in this comparison stem only from the differences in the methods, all hounds applied the same relative weight and surface correction scheme (where applicable) and the star was chosen among the ones having the same physics as the grid. We did not include the results from the methods SB and IR$_{\nu}$ in the comparison because they do not differ in a fundamental way from the method employed by VA. }

{The top panels of Fig.~\ref{fig:pdfs} show in black the results from the method employed by the hound JO. The MC simulations used in this method ensure the smoothness of the distributions for the inferred stellar properties. The uncertainties in this case are smaller than the uncertainties in the properties inferred by all other hounds, most noticeable for the age where the next smallest uncertainty (DR) is a factor of $\sim$3 larger.  Part of the reason could be that the perturbations to the individual frequencies were not considered in the MC simulations, to avoid the associated increase in computational time. To verify the impact of this approximation, a new MC simulation was performed, by decreasing the number of realisations but including perturbations to the individual frequencies. The results are highlighted in green in the same panels and show no significant change in the mass and radius distributions. However, the distribution for the age is found to be wider, with an associated uncertainty in age 1.7 times larger than in the case shown in black. Given the computational time involved in the MC simulations when the individual frequencies are perturbed, we can conclude that while this method  may be appropriate to model individual stars, it is not sufficiently efficient to be considered for a pipeline aimed at processing the data collected on many {thousands of} stars. }
%The method adopted by the hound JO (top panels) differs from all other methods in that the distributions of the stellar properties are derived following a series of Monte Carlo (MC) simulations, where the classical and global seismic constraints are perturbed by a normally-distributed random amount with variance given by the observational errors.  The MC simulations ensure the smoothness of the distributions for the inferred stellar properties. In the way the method is implemented here, the cost function includes global seismic constraints and the phase differences $\epsilon_l$ in the manner described by \cite{roxburgh16}, in addition to the classical constraints and the surface-corrected individual frequencies.  The uncertainties in the inferred properties are smaller than the uncertainties in the properties inferred by all other hounds (by factors of $\sim$1.5-3.5), but they still contain the true properties of Zebedee within 1$\sigma$.}

{The second row in Fig.~\ref{fig:pdfs} shows the results from the method adopted by the hound DR. This method  is unique in its sampling strategy, employing an MCMC approach coupled with interpolation on the grid. This approach results in distributions for the stellar properties that are also relatively smooth.  The uncertainties are only slightly smaller than those found  by the hound VA (fourth row) using the same set of constraints, without interpolation or the MCMC scheme. The main difference between the results of these two methods is in the smoothness of the distributions, with the probability distributions by VA showing significantly more structure.} 

{The third row in Fig.~\ref{fig:pdfs} shows the results from one of the methods adopted by the hound IR (IR$_\epsilon$), the only surface-independent method discussed in this work. Just as in the case of VA (fourth row), the IR$_\epsilon$ method does not perform grid interpolation, nor MCMC sampling. Therefore, the differences seen in the distributions of the properties inferred by these two methods likely follow mostly from the differences in the way the seismic data is used to constrain the models.  In fact, the surface-independent method IR$_\epsilon$ has significant constraining power on the age and a small constraining power on the radius, which is constrained mostly by the classical parameters and the frequency of the lowest frequency mode, as described before. In contrast, the individual frequencies fitted in the VA method have significant constraining power on all three stellar properties. }

{Finally, the last two rows in Fig.~\ref{fig:pdfs} compare the results from the method adopted by the hound VA and a variant of it VA$_{\rm int}$, including grid interpolation. When interpolation is considered, the probability distributions become smoother and their bimodal shape tends to disappear. The uncertainties also decrease somewhat, becoming closer to those derived by the hound DR.}

\section{Results for the elected methods}
\label{res}
%\red{[To solve: in the tables the relative difference to the true values is given in percentage. Shall we change the plots to show the Y axis also in percentage ? Then we could say in the captions of the figures that we are plotting $d_{\rm rel}^i (\%)$, perhaps making the correspondence to the tables easier for the reader. ]}

This section compares the results from the five methods elected for comparison, considering the six targets simulated for our hare-and-hounds exercise. The accuracy and precision of our grid-based inferences, as well as the biases detected when considering the results from all hounds will be discussed in Section~\ref{sec:comparison}, while the origin of the most significant discrepancies will be assessed in Section~\ref{sec: discrepancies}.

To quantify the accuracy, precision, and bias, we define a set of quantities and averages, as described below.
The accuracy of the inferred properties is determined by comparing them with the true, known values.
For each case $i$ ({i.e.} fixed {target} and method), we thus define measures of the relative and normalised differences to the truth, respectively,
\begin{equation}
	\label{eq:dif_rel}
	d_{\rm rel}^i = \frac{p^{\rm fit}_i-p^{\rm exact}}{p^{\rm exact}}\equiv \left(\frac{\delta p}{p}\right)_i,
\end{equation}
and
\begin{equation}
	\label{eq:dif_norm}
	d_{\rm norm}^i = \frac{p^{\rm fit}_i-p^{\rm exact}}{\sigma^{\rm fit}_i},
\end{equation}
where $p^{\rm fit}_i$ represents a stellar property inferred from a given fit, $\sigma^{\rm fit}_i$ the associated uncertainty and $p^{\rm exact}$ the corresponding true value. The notation $\delta p/p$, introduced in Eq.~(\ref{eq:dif_rel}), will be used in Figs~\ref{fig:comparison}-\ref{fig:errors}.  {Ideally, one would wish  $\left\lvert d_{\rm norm}^i\right\rvert$ to be smaller than one in $\sim$ 68~per~cent of the cases} and $\left\lvert d_{\rm rel}^i\right\rvert$ to be smaller than the accuracy requirement on the inference. 

Moreover, following \cite{reese16}, we define the average relative and normalised errors, respectively
\begin{equation}
	\label{eq:err_rel}
	\varepsilon_{\rm rel} = \sqrt{\frac{1}{N}\sum_i\left(d_{\rm rel}^i\right)^2}, 
\end{equation}
and
\begin{equation}
	\label{eq:err_norm}
	\varepsilon_{\rm norm} = \sqrt{\frac{1}{N}\sum_i\left(d_{\rm norm}^i\right)^2}.
\end{equation}
where the sum is taken over all {targets}, for a fixed method, or over all methods, for a fixed {target}, depending on the case considered. 

Following the same authors, the relative and normalised biases are measured, respectively, through
\begin{equation}
	\label{eq:bias_rel}
		b_{\rm rel} = {\frac{1}{N}\sum_ i d_{\rm rel}^i}, 
\end{equation}
and
\begin{equation}
	\label{eq:bias_norm}
	b_{\rm norm}= {\frac{1}{N}\sum_i d_{\rm norm}^i},
\end{equation}
with the sum taken over the {targets} or the methods, as above. 

Finally, the precision on a given property, in a given case is considered in relative terms through, 
\begin{equation}
\sigma_{\rm rel}^i=\frac{\sigma_i^{\rm fit}}{p^{\rm exact}},
\end{equation}
representing the 1$\sigma$ error bar on the quantity $d_{\rm rel}^i$. The average precision for a given {target} or method, $\sigma_{\rm rel}$, is obtained by averaging $\sigma_{\rm rel}^i$ over all methods or targets, respectively. A larger value of $\sigma_{\rm rel}$ implies a less precise inference. 

{In the computation of the quantities defined above, the values of the inferred properties were taken to be the means and standard deviations derived from the corresponding probability distributions. Comparison of the means and the 50th percentiles show that they provide very close point estimates for the stellar properties. In most cases, the difference between the two does not exceed 0.2$\sigma$, with only a few cases reaching 0.6$\sigma$. Only in the case of the target George, the difference was found to be yet larger, for one of the hounds.}

As guidance, in Section~\ref{sec:comparison} we compare the relative quantities (differences, error, biases, and precision) with the accuracy requirements set by PLATO for a G0V star of magnitude $V=10$, respectively, 15, 2, and 10 per cent on stellar mass, radius, and age ({\it hereafter}, the reference values).\footnote{ESA PLATO Science Requirements Document (PTO-EST-SCI-RS-0150\_SciRD\_8\_0)} These follow from the requirements set on the mass, radius, and age determination of the exoplanets to be characterised by the mission \citep{rauer14}.\footnote{https://sci.esa.int/web/plato/-/42277-science}
%\citep{rauer14}. 
%\red{[Marie Jo: should we cite the ESA SciRD document here? (I don't think the reference star for these requirements is given in the Rauer+2014) How do we cite that?]}
%%% DRR: I hadn't realised that the error tolerance on the mass was now 15% - this seems very large

\subsection{Accuracy and precision of the elected methods}
\label{sec:comparison}

\begin{figure*}
	\includegraphics[width=2.02\columnwidth]{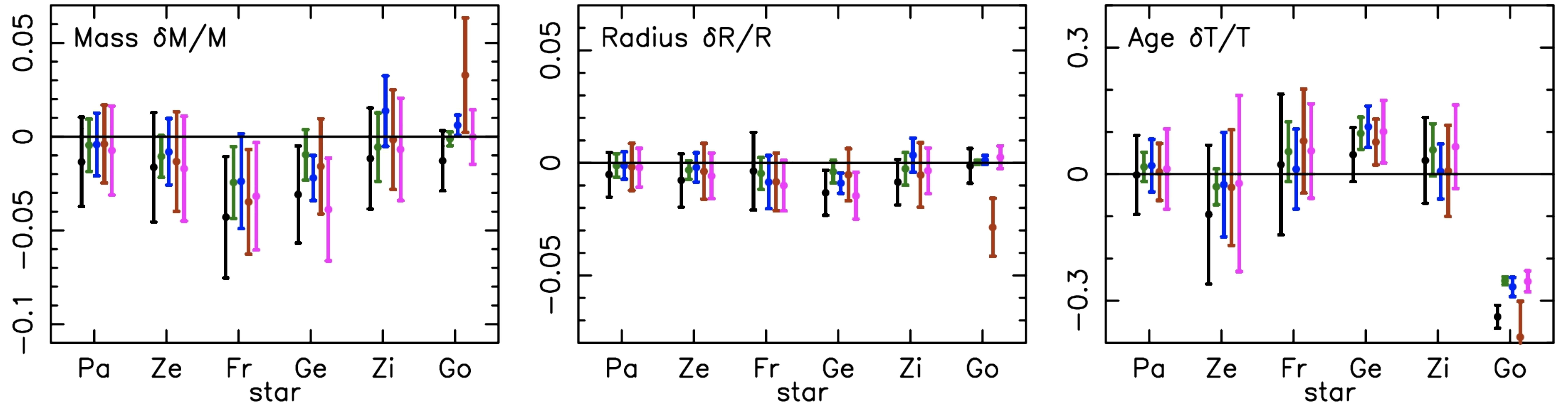}
    \caption{Relative differences between the stellar properties inferred for each {target} and the corresponding true values (as defined in Eq.~(\ref{eq:dif_rel})). For the corresponding values, expressed as a percentage, see $d_{\rm rel}^i$ in tables~\ref{tab:mass} to \ref{tab:age}. Left panel: relative mass difference. Middle panel: relative radius difference. Right panel: relative age differences. {Targets} are identified according to their ID and hounds according to their colour, listed in Tables~\ref{tab:hares} and \ref{tab:hounds}, respectively. For each {target}, 5 inferences are shown, corresponding to the results elected {for} the 5 hounds.}
    \label{fig:comparison}
\end{figure*}

%%% DRR: I replaced "by" by "for", as it wasn't the hounds themselves who elected the method but rather Margarida and Ian.
The mass, radius and age inferred for the six {targets} using the method elected {for} each hound are shown in Fig.~\ref{fig:comparison} and a summary of the corresponding results is given in Tables~\ref{tab:mass}-\ref{tab:age}. Most hounds reported having a problem when attempting to fit George, suggesting that the {target} falls outside the parameter space covered by the grid. This example shows how problems with the grid can be flagged based on the solutions found, at least when no alternative (degenerate) good solutions exist within the parameter space. For completeness, we report the results from fitting George in Fig.~\ref{fig:comparison} and Tables~\ref{tab:mass}-\ref{tab:age} but do not consider them in the computation of the bias, average errors, and average precision for each hound (last three columns in Tables~\ref{tab:mass}-\ref{tab:age}) and will also disregard them in the analysis of results that follows below. For the remaining five {targets}, the accuracy and precision of the inferred mass, radius and age are, with one single exception,  within the reference values.

For the mass, the {most significant relative difference}, $\max\left(\left|{d_{\rm rel}^i}\right|\right)$, is found for Fred and amounts to $4.32$~per~cent. This is well within the reference value of $15$~per~cent for stellar mass. Fred is also the {target} showing the highest average relative error (3.24 per cent) and the {most significant} relative bias (-3.16 per cent) on mass. In fact, an inspection of Fig.~\ref{fig:comparison} and Tables~\ref{tab:mass}-\ref{tab:age} shows that all hounds inferred a mass slightly smaller (between $\sim$ 2--4 per cent) than the true mass for this {target}. Moreover, in most cases the inferred value for the mass of Fred is slightly more than 1\,$\sigma$ away from the true value, resulting in a normalised average error of 1.19. The next most significant mass discrepancy is found for Gerald, with an average relative error on the mass of only 2.58 per cent and an average normalised error of 1.25. Also for this {target}, the mass has been systematically underestimated, with a resulting relative bias of -2.36 per cent.

For the radius, the {most significant relative difference} is found for Gerald, amounting to $1.33$~per~cent (to be compared with the reference value of 2~per~cent). Nevertheless, the average relative error for Gerald is only 0.96 per cent, reflecting that most hounds found a relative difference in radius whose magnitude is below 1 per cent for this {target}. On the other hand, the radius normalised average error is 1.27 for Gerald, indicating that for some hounds the inferred radius is more than 1\,$\sigma$ away from the true radius of this {target}, as can be confirmed through inspection of Fig.~\ref{fig:comparison}. Nevertheless, Gerald is a true exception. For the other four {targets}, and for all five hounds, the magnitude of the relative difference in radius is  below 1.02 per cent and the magnitude of the normalised differences is  below 1. 

For the age, the {most significant relative difference} is again found for Gerald, amounting to 11.25 per cent. This is the only {target} (George excluded) for which the absolute value of the relative difference in age found by one of the hounds exceeds the 10 per cent reference value. Nevertheless, the relative differences found by the other four hounds for the age of Gerald {are below 10 per cent}, the final average relative error being 8.96 per cent. Gerald is also the only {target} with an average normalised age error larger than 1, reflecting the fact that the age inferred by most hounds differs more than 1\,$\sigma$ from the true value.   Also worth noting is the relative bias in the age inferred for this {target} (8.66 per cent), with all hounds overestimating its age. This is not surprising, given the bias towards lower masses found in the results for Gerald, as discussed above.  Somewhat significant biases in age, of -4.06~per~cent and 4.43~per~cent, are also found for Zebedee and Fred, respectively. However, in these cases, the true age values are  within 1\,$\sigma$ of the ages inferred by all hounds.

Finally, the results presented in Tables~\ref{tab:mass}-\ref{tab:age} show that the average precision for each {target}, $\sigma_{\rm rel}$, is typically within twice the average relative error. The most notable exception is Patch for which we find that the $\sigma_{\rm rel}$ values for the radius and age are approximately 3 and 5 times larger than the corresponding values of $\varepsilon_{\rm rel}$, respectively. {Moreover, for the ages of Zebedee and Fred, we note that four and three  out of the five hounds, respectively, find a $\sigma_{\rm rel}^i$ larger than 10 per cent. Nevertheless, neither of these stars is a good representative of the PLATO reference star. Zebedee, while having a mass of $\sim$1~M$_\odot$, is  {much} younger than our sun, and Fred, with a mass of $\sim$1.4~M$_\odot$, is significantly more massive.}

\subsection{Origin of the most significant discrepancies}
\label{sec: discrepancies}
%%% DRR: please choose between notion and idea (don't keep both)
{The results reported in Section~\ref{sec:comparison} support {the idea} that grid-based inference procedures are a viable option to infer accurate stellar properties of main-sequence stars, as required by PLATO.} Still, it is of interest to understand the origin of the systematic differences found in the results for some of the {targets} considered in this exercise. That understanding is important both to anticipate the systematic errors that may be present in the analysis of real PLATO data and to help design optimal grids for the grid-based inference procedure that will be adopted. Among the six {targets} that were modelled, three have proven to be more challenging, with the inferred properties being systematically off and/or more than 1\,$\sigma$ away from the known true values. In what follows we discuss the physical origin of these differences.\\

\noindent{\bf George:} As mentioned in Section~\ref{sec:comparison}, most hounds reported not having been able to find an adequate model for George within the parameter space covered by the {provided} grid. This is the optimal report in the case of George, since the overshoot adopted for this {target} is, indeed, significantly larger than the values considered when constructing the grid  (cf. Section~\ref{hares}). The fact that the hounds were able to identify the problem shows that no other combination of parameters within the grid could mimic the observational data for George. Unfortunately, that is not  the case, as we shall see from the discussion {for} Fred below. In the case of George, the inadequacy of the grid concerning the range of overshoot has a significant impact on the inferred age, which is found to be smaller than the true age in all cases (Fig.~\ref{fig:comparison}, right panel). In addition, the inferred initial helium mass fraction for George is found { to be} significantly larger than that {used} when generating the {target}.\\ 

\noindent{\bf Fred:} The mass inferred for Fred was systematically smaller than the true mass, while its inferred age was found to be systematically larger than the true age. Given that the physics adopted to generate this {target} was the same as the physics adopted to construct the grid, the origin of these differences is expected to be in the limits of the parameter space covered by the grid. The mass of the {target}, $M=1.4318$~M$_{\odot}$, is relatively close to the upper limit of the mass in the grid (0.5~$\le M_{\rm grid} \le 1.5$~M$_{\odot}$). This could impact the tail of the inferred mass probability density function and, thus, bias the inferred mass. However, inspection of the mass probability density functions inferred by the hounds shows that the tails of the mass distributions are well within the grid mass limits. Alternatively, the discrepancy could stem from the chemical composition of the {target}, in particular, from the relation between the initial helium and metal mass fractions. In fact, considering the values of $Y_{\rm ini}$ and $Z_{\rm ini}$ used to generate Fred (Table~\ref{tab:hares}) and the {Big Bang nucleosynthesis} helium mass fraction adopted in the grid, $Y_0=0.248$, we find an enrichment ratio for Fred of $dY/dZ$=0.77, thus, smaller than the lower limit of the grid ($1\le {d}Y/{d}Z_{\rm grid}\le 3$). Therefore, for a given $Z_{\rm ini}$, the models will have a larger $Y_{\rm ini}$, hence a larger mean molecular weight which, at fixed mass, would lead to an increase in central temperature and, thus, in luminosity. Since both the luminosity and metallicity are constrained, the best solution is found, instead, for models with a lower mass. This near-degeneracy between stellar mass and initial helium mass fraction is well known \cite[e.g.][]{cunha03,lebreton14,nsamba21} and in the current case prevented the hounds from detecting that the grid was not adequate because its parameter space did not cover the value of the enrichment ratio required to model Fred. {Additional information on the helium abundance, such as that contained in the seismic signature of the helium glitch, can help lift this degeneracy \citep{gough90,verma17,cunha20}. In particular, the characterisation of the helium glitch signature and its use in the grid-based inference, could be key in cases like the one discussed here. These results can thus be useful while designing the PLATO stellar pipeline, as well as when deciding on the characteristics of the grid that will be associated with it, particularly when considering whether to {rely on} an enrichment law or to let $Y_{\rm ini}$ and $Z_{\rm ini}$ vary freely in the grid.} \\

\noindent{\bf Gerald:} As in the case of Fred, the mass inferred for Gerald was found to be systematically smaller than the true mass, while its inferred age was found to be systematically larger than the true age. However, in the case of Gerald, the origin of the discrepancy lies in the physics adopted to generate the {target}, which, unlike in the case of the grid, included atomic diffusion. The inclusion of diffusion in the {target} results in an observed surface iron abundance [Fe/H]$_{\rm obs}$ smaller than the initial value. The models in the grid do not incorporate that evolutionary change in the surface abundance of iron, thus the constraint on [Fe/H]$_{\rm obs}$ imposed when fitting the models to the observations, bias the models towards an initial surface iron abundance that is smaller than the one used to generate the {target}. This could be achieved in two ways, namely, a decrease in $Z_{\rm ini}$ and/or a decrease in $Y_{\rm ini}$ (implying an increase in $X_{\rm ini})$. Inspection of the solutions reveals that the main impact, in this case, results from the metallicity. In fact, all hounds found a $Z_{\rm ini}$ lower than the true value. The lower $Z_{\rm ini}$ implies a lower opacity in the core, hence the potential for an increase in energy transport. To avoid the consequent increase in luminosity, which is constrained by the observations, the best solutions have a mass that is lower than the true value. This near degeneracy between metallicity and mass and its impact on the inferred stellar age is also well known and reported in the literature  \cite[e.g.][]{cunha03,nsamba18}.

\begin{table*}
	\centering
	\caption{True and inferred stellar masses for the six {targets}, all given in units of the solar mass. Also shown are the relative $d_{\rm rel}^i$ and normalised $d_{\rm norm}^i$ differences (cf. Eqs~(\ref{eq:dif_rel}) and (\ref{eq:dif_norm}), respectively). For each hound, the relative and normalised biases (computed considering all {targets}, except George) are shown in the 9$^{\rm th}$ column, the average relative $\varepsilon_{\rm rel}$ and average normalised $\varepsilon_{\rm norm}$ errors are shown in the 10$^{\rm th}$ column, and the average precision is given the the 11$^{\rm th}$ (last) column. The last five rows show, for each {target}, the biases, average errors, and average precision, considering the results from all hounds.}
	\label{tab:mass}
	\resizebox{2\columnwidth}{!}{%
	\begin{tabular}{llccccccrrr} % four columns, alignment for each
		\hline
&	{\bf Hares}	&	Patch	&	Zebedee	&	Fred	&	Gerald	&	Zippy	&	George	&		&		&		\\
&	Mass	&	0.8644	&	1.0165	&	1.4318	&	1.0242	&	1.1278	&	1.3430	&		&		&		\\
\hline																				
&		&		&		&		&		&		&		&	$b_{\rm rel}$ ($\%$)*	&	$\varepsilon_{\rm rel}$ ($\%$)*	&	$\sigma_{\rm rel}$ ($\%$)*	\\
{\bf Hounds} &		&		&		&		&		&		&		&	$b_{\rm norm}$ *	&	$\varepsilon_{\rm norm}$*	&		\\
\hline																				
&	Mass	&	0.853(0.020)	&	1.000(0.030)	&	1.370(0.046)	&	0.992(0.027)	&	1.115(0.030)	&	1.326(0.022)	&		&		&		\\
SB &	$d_{\rm rel}^i$  ($\%$)	&	-1.32	&	-1.62	&	-4.32	&	-3.14	&	-1.13	&	-1.27	&	-2.31	&	2.61	&		\\
&	$d_{\rm norm}^i$ 	&	-0.57	&	-0.55	&	-1.34	&	-1.19	&	-0.43	&	-0.77	&	-0.82	&	0.90	&		\\
&	$\sigma_{\rm rel}^i$  ($\%$)	&	2.31	&	2.95	&	3.21	&	2.64	&	2.66	&	1.64	&		&		&	2.75	\\
\hline																				
&	Mass	&	0.860(0.012)	&	1.006(0.011)	&	1.397(0.027)	&	1.014(0.014)	&	1.121(0.021)	&	1.3414(0.0051)	&		&		&		\\
JO &	$d_{\rm rel}^i$  ($\%$)	&	-0.46	&	-1.06	&	-2.44	&	-0.96	&	-0.56	&	-0.12	&	-1.10	&	1.31	&		\\
&	$d_{\rm norm}^i$ 	&	-0.33	&	-0.95	&	-1.27	&	-0.71	&	-0.31	&	-0.31	&	-0.71	&	0.80	&		\\
&	$\sigma_{\rm rel}^i$  ($\%$)	&	1.39	&	1.11	&	1.92	&	1.35	&	1.83	&	0.37	&		&		&	1.52	\\
\hline																				
&	Mass	&	0.861(0.014)	&	1.008(0.018)	&	1.398(0.036)	&	1.002(0.012)	&	1.143(0.021)	&	1.3512(0.0073)	&		&		&		\\
DR &	$d_{\rm rel}^i$  ($\%$)	&	-0.39	&	-0.84	&	-2.36	&	-2.17	&	1.37	&	0.61	&	-0.88	&	1.61	&		\\
&	$d_{\rm norm}^i$ 	&	-0.24	&	-0.47	&	-0.94	&	-1.85	&	0.73	&	1.12	&	-0.55	&	1.01	&		\\
&	$\sigma_{\rm rel}^i$  ($\%$)	&	1.62	&	1.77	&	2.51	&	1.17	&	1.86	&	0.54	&		&		&	1.79	\\
\hline																				
&	Mass	&	0.861(0.018)	&	1.003(0.027)	&	1.382(0.040)	&	1.008(0.026)	&	1.126(0.030)	&	1.387(0.041)	&		&		&		\\
IR$_\epsilon$  &	$d_{\rm rel}^i$  ($\%$)	&	-0.39	&	-1.33	&	-3.48	&	-1.58	&	-0.16	&	3.28	&	-1.39	&	1.82	&		\\
&	$d_{\rm norm}^i$ 	&	-0.19	&	-0.50	&	-1.25	&	-0.62	&	-0.06	&	1.07	&	-0.52	&	0.67	&		\\
&	$\sigma_{\rm rel}^i$  ($\%$)	&	2.08	&	2.66	&	2.79	&	2.54	&	2.66	&	3.05	&		&		&	2.55	\\
\hline																				
&	Mass	&	0.858(0.020)	&	0.999(0.028)	&	1.386(0.041)	&	0.984(0.028)	&	1.120(0.031)	&	1.343(0.019)	&		&		&		\\
VA &	$d_{\rm rel}^i$  ($\%$)	&	-0.74	&	-1.72	&	-3.20	&	-3.93	&	-0.69	&	0.00	&	-2.06	&	2.43	&		\\
&	$d_{\rm norm}^i$ 	&	-0.31	&	-0.62	&	-1.12	&	-1.43	&	-0.25	&	0.00	&	-0.75	&	0.88	&		\\
&	$\sigma_{\rm rel}^i$  ($\%$)	&	2.37	&	2.79	&	2.86	&	2.74	&	2.73	&	1.45	&		&		&	2.70	\\
\hline																				
&	$b_{\rm rel}$ ($\%$)	&	-0.66	&	-1.31	&	-3.16	&	-2.36	&	-0.24	&	0.50	&	&			&		\\
&	$b_{\rm norm}$ 	&	-0.33	&	-0.62	&	-1.18	&	-1.16	&	-0.06	&	0.22	&	&			&		\\
&	$\varepsilon_{\rm rel}$ ($\%$)	&	0.75	&	1.36	&	3.24	&	2.58	&	0.89	&	1.60	&	&			&		\\
&	$\varepsilon_{\rm norm}$	&	0.35	&	0.64	&	1.19	&	1.25	&	0.42	&	0.79	&	&			&		\\
&	$\sigma_{\rm rel}$ ($\%$)	&	1.95	&	2.26	&	2.66	&	2.09	&	2.35	&	1.41	&	&		&			\\
\hline
 \multicolumn{10}{l}{*Averages performed excluding the results for George (see text for details).} \\
	\end{tabular}%
	}
\end{table*}
\begin{table*}
	\centering
	\caption{True  and inferred stellar radii for the six {targets}, all given in units of the solar radius. Rows and columns as in Table~\ref{tab:mass}.}
	\label{tab:radius}
	\resizebox{2\columnwidth}{!}{%
	\begin{tabular}{llccccccrrr} % four columns, alignment for each
		\hline
&	{\bf Hares}	&	Patch	&	Zebedee	&	Fred	&	Gerald	&	Zippy	&	George	&		&		&		\\
&	Radius	&	0.9557	&	0.9646	&	1.7225	&	1.2053	&	1.3965	&	1.7069	&		&		&		\\
\hline																				
&		&		&		&		&		&		&		&	$b_{\rm rel}$ ($\%$)*	&	$\varepsilon_{\rm rel}$ ($\%$)*	&	$\sigma_{\rm rel}$ ($\%$)*	\\
{\bf Hounds} &		&		&		&		&		&		&		&	$b_{\rm norm}$ *	&	$\varepsilon_{\rm norm}$*	&		\\
\hline																				
&	Radius	&	0.9507(0.0095)	&	0.957(0.011)	&	1.716(0.011)	&	1.189(0.012)	&	1.385(0.014)	&	1.705(0.013)	&		&		&		\\
SB &	$d_{\rm rel}^i$  ($\%$)	&	-0.52	&	-0.78	&	-0.36	&	-1.33	&	-0.85	&	-0.13	&	-0.77	&	0.84	&		\\
&	$d_{\rm norm}^i$ 	&	-0.52	&	-0.65	&	-0.21	&	-1.32	&	-0.84	&	-0.17	&	-0.71	&	0.80	&		\\
&	$\sigma_{\rm rel}^i$  ($\%$)	&	0.99	&	1.19	&	1.73	&	1.00	&	1.01	&	0.77	&		&		&	1.18	\\
\hline																				
&	Radius	&	0.9546(0.0050)	&	0.9616(0.0040)	&	1.714(0.012)	&	1.2006(0.0061)	&	1.393(0.010)	&	1.7075(0.0016)	&		&		&		\\
JO &	$d_{\rm rel}^i$  ($\%$)	&	-0.12	&	-0.31	&	-0.47	&	-0.39	&	-0.26	&	0.04	&	-0.31	&	0.33	&		\\
&	$d_{\rm norm}^i$ 	&	-0.22	&	-0.76	&	-0.66	&	-0.77	&	-0.36	&	0.40	&	-0.56	&	0.60	&		\\
&	$\sigma_{\rm rel}^i$  ($\%$)	&	0.53	&	0.41	&	0.71	&	0.50	&	0.72	&	0.09	&		&		&	0.58	\\
\hline																				
&	Radius	&	0.9546(0.0058)	&	0.9627(0.0063)	&	1.708(0.020)	&	1.1945(0.0055)	&	1.401(0.010)	&	1.7091(0.0035)	&		&		&		\\
DR &	$d_{\rm rel}^i$  ($\%$)	&	-0.12	&	-0.20	&	-0.86	&	-0.90	&	0.34	&	0.13	&	-0.35	&	0.58	&		\\
&	$d_{\rm norm}^i$ 	&	-0.19	&	-0.31	&	-0.72	&	-1.99	&	0.45	&	0.63	&	-0.55	&	0.98	&		\\
&	$\sigma_{\rm rel}^i$  ($\%$)	&	0.61	&	0.65	&	1.18	&	0.45	&	0.76	&	0.20	&		&		&	0.73	\\
\hline																				
&	Radius	&	0.954(0.010)	&	0.961(0.012)	&	1.708(0.022)	&	1.199(0.014)	&	1.389(0.020)	&	1.658(0.022)	&		&		&		\\
IR$_\epsilon$  &	$d_{\rm rel}^i$  ($\%$)	&	-0.18	&	-0.37	&	-0.84	&	-0.52	&	-0.54	&	-2.86	&	-0.49	&	0.54	&		\\
&	$d_{\rm norm}^i$ 	&	-0.17	&	-0.30	&	-0.66	&	-0.45	&	-0.38	&	-2.22	&	-0.39	&	0.42	&		\\
&	$\sigma_{\rm rel}^i$  ($\%$)	&	1.05	&	1.24	&	1.28	&	1.16	&	1.43	&	1.29	&		&		&	1.23	\\
\hline																				
&	Radius	&	0.954(0.008)	&	0.959(0.010)	&	1.705(0.019)	&	1.190(0.012)	&	1.392(0.014)	&	1.711(0.009)	&		&		&		\\
VA &	$d_{\rm rel}^i$  ($\%$)	&	-0.18	&	-0.58	&	-1.02	&	-1.27	&	-0.32	&	0.24	&	-0.67	&	0.79	&		\\
&	$d_{\rm norm}^i$ 	&	-0.21	&	-0.57	&	-0.90	&	-1.23	&	-0.32	&	0.48	&	-0.64	&	0.75	&		\\
&	$\sigma_{\rm rel}^i$  ($\%$)	&	0.86	&	1.01	&	1.13	&	1.03	&	1.01	&	0.50	&		&		&	1.01	\\
\hline																				
&	$b_{\rm rel}$ ($\%$)	&	-0.22	&	-0.45	&	-0.71	&	-0.88	&	-0.33	&	-0.52	&	&			&		\\
&	$b_{\rm norm}$ 	&	-0.26	&	-0.52	&	-0.63	&	-1.15	&	-0.29	&	-0.18	&	&			&		\\
&	$\varepsilon_{\rm rel}$ ($\%$)	&	0.27	&	0.49	&	0.75	&	0.96	&	0.51	&	1.29	&	&			&		\\
&	$\varepsilon_{\rm norm}$	&	0.29	&	0.55	&	0.67	&	1.27	&	0.51	&	1.07	&	&			&		\\
&	$\sigma_{\rm rel}$ ($\%$)	&	0.81	&	0.90	&	1.20	&	0.83	&	0.99	&	0.57	&	&			&		\\
\hline
 \multicolumn{11}{l}{*Averages performed excluding the results for George (see text for details).} \\
	\end{tabular}%
	}
\end{table*}

\begin{table*}
	\centering
	\caption{True and inferred stellar ages for the six {targets}, all given in Gyr. Rows and columns as in Table~\ref{tab:mass}.}
	\label{tab:age}
	\resizebox{2\columnwidth}{!}{%
	\begin{tabular}{llccccccrrr} % four columns, alignment for each
		\hline
&	{\bf Hares}	&	Patch	&	Zebedee	&	Fred	&	Gerald	&	Zippy	&	George	&		&		&		\\
&	Age	&	9.898	&	3.085	&	1.839	&	8.039	&	4.223	&	3.757	&		&		&		\\
\hline																				
&		&		&		&		&		&		&		&	$b_{\rm rel}$ ($\%$)*	&	$\varepsilon_{\rm rel}$ ($\%$)*	&	$\sigma_{\rm rel}$ ($\%$)*	\\
{\bf Hounds} &		&		&		&		&		&		&		&	$b_{\rm norm}$ *	&	$\varepsilon_{\rm norm}$*	&		\\
\hline																				
&	Age	&	9.88(0.93)	&	2.79(0.51)	&	1.88(0.31)	&	8.41(0.52)	&	4.36(0.43)	&	2.49(0.10)	&		&		&		\\
SB &	$d_{\rm rel}^i$  ($\%$)	&	-0.19	&	-9.56	&	2.28	&	4.62	&	3.27	&	-33.78	&	0.08 	&	5.07	&		\\
&	$d_{\rm norm}^i$ 	&	-0.02	&	-0.58	&	0.14	&	0.72	&	0.32	&	-12.46	&	0.12	&	0.44	&		\\
&	$\sigma_{\rm rel}^i$  ($\%$)	&	9.37	&	16.46	&	16.68	&	6.41	&	10.18	&	2.71	&		&		&	11.82	\\
\hline																				
&	Age	&	10.07(0.34)	&	2.99(0.13)	&	1.94(0.13)	&	8.82(0.31)	&	4.47(0.26)	&	2.807(0.035)	&		&		&		\\
JO &	$d_{\rm rel}^i$  ($\%$)	&	1.72	&	-3.00	&	5.34	&	9.70	&	5.75	&	-25.30	&	3.90	&	5.79	&		\\
&	$d_{\rm norm}^i$ 	&	0.50	&	-0.71	&	0.75	&	2.55	&	0.93	&	-27.20	&	0.80	&	1.32	&		\\
&	$\sigma_{\rm rel}^i$  ($\%$)	&	3.45	&	4.24	&	7.12	&	3.80	&	6.20	&	0.93	&		&		&	4.96	\\
\hline																				
&	Age	&	10.10(0.62)	&	3.01(0.38)	&	1.86(0.17)	&	8.94(0.40)	&	4.25(0.28)	&	2.753(0.085)	&		&		&		\\
DR &	$d_{\rm rel}^i$  ($\%$)	&	2.03	&	-2.46	&	1.22	&	11.25	&	0.64	&	-26.72	&	2.53	&	5.26	&		\\
&	$d_{\rm norm}^i$ 	&	0.32	&	-0.20	&	0.13	&	2.29	&	0.10	&	-11.78	&	0.53	&	1.04	&		\\
&	$\sigma_{\rm rel}^i$  ($\%$)	&	6.26	&	12.38	&	9.46	&	4.92	&	6.51	&	2.27	&		&		&	7.90	\\
\hline																				
&	Age	&	9.95(0.67)	&	2.99(0.42)	&	1.98(0.23)	&	8.65(0.44)	&	4.26(0.46)	&	2.31(0.32)	&		&		&		\\
IR$_\epsilon$  &	$d_{\rm rel}^i$  ($\%$)	&	0.52	&	-3.18	&	7.83	&	7.63	&	0.76	&	-38.62	&	2.71	&	5.11	&		\\
&	$d_{\rm norm}^i$ 	&	0.08	&	-0.23	&	0.63	&	1.41	&	0.07	&	-4.56	&	0.39	&	0.70	&		\\
&	$\sigma_{\rm rel}^i$  ($\%$)	&	6.74	&	13.71	&	12.34	&	5.42	&	10.82	&	8.46	&		&		&	9.81	\\
\hline																				
&	Age	&	10.019(0.94)	&	3.02(0.64)	&	1.94(0.21)	&	8.85(0.59)	&	4.50(0.42)	&	2.80(0.09)	&		&		&		\\
VA &	$d_{\rm rel}^i$  ($\%$)	&	1.22	&	-2.11	&	5.49	&	10.09	&	6.56	&	-25.47	&	4.25	&	6.01	&		\\
&	$d_{\rm norm}^i$ 	&	0.13	&	-0.10	&	0.49	&	1.36	&	0.66	&	-10.30	&	0.51	&	0.72	&		\\
&	$\sigma_{\rm rel}^i$  ($\%$)	&	9.52	&	20.86	&	11.18	&	7.39	&	9.89	&	2.47	&		&		&	11.77	\\
\hline																				
&	$b_{\rm rel}$ ($\%$)	&	1.06	&	-4.06	&	4.43	&	8.66	&	3.40	&	-30.0	&	&			&		\\
&	$b_{\rm norm}$ 	&	0.20	&	-0.36	&	0.43	&	1.67	&	0.42	&	-13.3	&	&			&		\\
&	$\varepsilon_{\rm rel}$ ($\%$)	&	1.33	&	4.92	&	5.03	&	8.96	&	4.19	&	30.4	&	&			&		\\
&	$\varepsilon_{\rm norm}$	&	0.27	&	0.43	&	0.50	&	1.79	&	0.53	&	15.2	&	&			&		\\
&	$\sigma_{\rm rel}$ ($\%$)	&	7.07	&	13.53	&	11.36	&	5.59	&	8.72	&	3.37	&	&			&			\\
\hline
 \multicolumn{10}{l}{*Averages performed excluding the results for George (see text for details).} \\
	\end{tabular}%
	}
\end{table*}

\section{{Impact from surface corrections}}
\label{sec:surf}
{When fitting individual frequencies, the use of an inadequate empirical correction to account for the systematic offsets in the model frequencies can introduce biases in the inferred stellar properties. Unfortunately, as our data are obtained from simulations that are themselves based on stellar models, the surface effects incorporated in the "observed" oscillation frequencies are also derived from an empirical prescription and may not capture the truth that we would like to simulate. While this limits our ability to quantify the impact on the inferred stellar properties from the true, unknown, surface effects, one can at the least quantify the impact of  correcting the model frequencies with an empirical correction that differs from the one employed in the simulations, as well as the impact of not correcting the model frequencies at all. We recall that the one-term correction by \cite{ball14} was used to mimic the surface effects in the simulations of the seismic data (see Section~\ref{hares} for details).  }

\begin{figure*}
	\includegraphics[width=2\columnwidth]{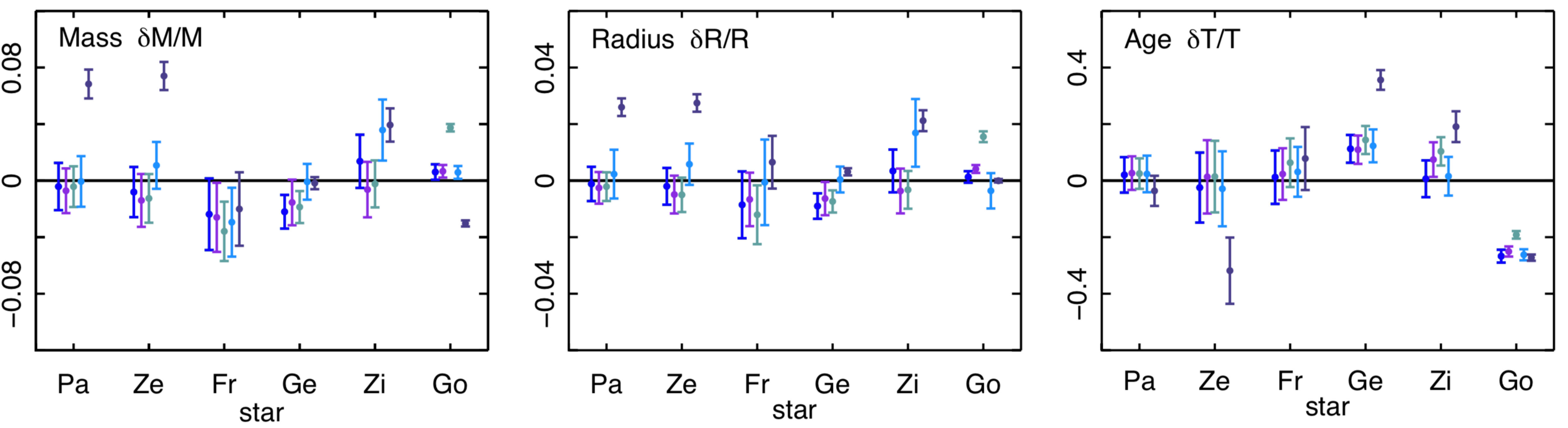}
    \caption{Inferences for a single hound (DR) when considering different prescriptions for the surface corrections. Panels as in Fig.~\ref{fig:comparison}. For each target, five inferences are shown in the following order, from left to right: 1. \protect\cite{ball14} two-term correction (blue), 2. \protect\cite{ball14} one-term correction (purple), 3. \protect\cite{kjeldsen08} (cadet), 4. \protect\cite{sonoi15} (light blue), 5. No correction (dark blue). }
    \label{fig:surface}
\end{figure*}

{Figure~\ref{fig:surface} illustrates the impact on the inferred stellar properties from employing different formulations of the surface corrections published in the literature. The inferences shown were performed using the method by DR, with a 3:3 weight. Considering the four inferences performed using some form of empirical corrections (first four results for each target in Fig.~\ref{fig:surface}), and excluding George, the most significant relative differences {with the true values} are found to be 3.6 per cent for mass (for Zippy and Fred), 1.7 per cent for radius (for Zippy), and 14 per cent for age (for Gerald). {The latter is an example of how  a result comparable with the reference value, such as the age inference reported with {a given} method by DR in Table~\ref{tab:age} (11.25~per~cent), can become significantly larger than the reference when a different surface correction is considered.} More significant differences are found when no surface  {corrections are} applied, namely 7.4 per cent for mass and 2.7 per cent for radius (Zebedee), and 36 per cent for age (Gerald). Interestingly, the results do not seem to depend very significantly on the form of the surface correction adopted. We can quantify that dependence by taking as a reference the relative difference {with} the true value obtained with the elected method {for} DR, ($d_{\rm rel}^{\rm ref}$; blue in Fig.~\ref{fig:surface}), and computing the dispersion of the results for each target as $\sqrt{\sum_i(d_{\rm rel}^i-d_{\rm rel}^{\rm ref})^2/N_{\rm c}}$, where $N_{\rm c}=3$ is the number of cases considered for comparison, where we exclude the case with no surface correction. We find a maximum dispersion of 1.9 per cent for mass, 1.0 per cent for radius, and 6.8 per cent for age, all for the same target (Zippy).   }

\section{{Impact from {applying different weight schemes}}}
\label{weight}
%\red{[Victor: we need to discuss the BASTA results in Fig. 2 and also whether to include BASTA results in Fig. 4.]}\\
%the 3:1 from BASTA should be the same as in Fig. 1. Also, it is slightly odd that for Zebedee the 3:3 case of VA for mass and radius has larger error bars than the 3:1.
%Also, it would be interesting to add at least for the two IR cases, the result of 3:0 to figure 2, to have an idea of the improvement brought by the seismic data.
%]}\\

%In this section we discuss the impact on the inferred stellar properties from changing the weights given to the classical and seismic constraints when performing the forward modelling.
{When models provide a faithful representation of the truth and the differences between model predictions and observations result solely from measurement errors, one may confidently determine the uncertainty in the inferences {that are} made (often called internal or formal errors) by propagating the measurement uncertainties. In the context of this study, the formal errors are those inferred with a weight of 3:N, meaning that each observation is given the same weight in the likelihood function.  Unfortunately, perfect models are often not available and one is faced with having to {also} consider differences between model predictions and observations that may come about due to the improper modelling of the stars. {One way to tackle this problem consists in making a number of inferences based on different model sets, computed with different physics. The dispersion of the inferences is then either provided separately or added in quadrature to the formal error derived for one particular inference. }
%However, that approach does not address two  problems that are specific  {to} the asteroseismic modelling of stars: 1) that some of the differences between the model and observed frequencies result from an improper modelling that cannot be bracketed by varying the physics adopted in the model computation ({\it e.g.} the surface effects discussed in Section~\ref{sec:surf}) and 2) that the errors on the model frequencies are {sometimes much larger than the uncertainties in the measured frequencies}. In particular, the latter may result in the likelihood becoming very sensitive to the (inaccurate) individual mode frequency predictions, with the global constraints hardly influencing the final inference in those cases. To deal with this potential problem, it has become relatively common practice to {introduce} weights when defining the likelihood function.   }
{However, that approach does not address a problem that is specific  {to} the asteroseismic modelling of stars, namely, that some of the differences between the model and observed frequencies result from an improper modelling that cannot be bracketed by varying the physics adopted in the model computation ({\it e.g.} the surface effects discussed in Section~\ref{sec:surf}) and that the consequent errors on the model frequencies are {sometimes much larger than the uncertainties in the measured frequencies}. This may result} in the likelihood becoming very sensitive to the (inaccurate) individual mode frequency predictions, with the global constraints hardly influencing the final inference in those cases. To deal with this potential problem, it has become relatively common practice to {introduce} weights when defining the likelihood function.   }
%%% DRR: personally, I would say that the model errors on the frequencies are much larger than the uncertaintites in the measured frequencies.

{It is worth noting that the application of relative weights in the construction of the likelihood function is essentially equivalent to inflating the errors in the observed frequencies. {In fact,
% if the uncertainties in all frequencies were the same, 
applying a weighting scheme of 3:1 is equivalent to inflating the errors on the observed frequencies by a factor of $\sqrt N$, when computing the $\chi^2$ function. Likewise, the 3:3 case corresponds to a frequency error inflation of $\sqrt{N /3}$, and the 3:N case to taking the errors on the frequencies at face value.} Only the last case has a clear statistical interpretation, with the resulting uncertainties in the inferred values corresponding to the formal errors. It is, thus, important to understand how the application of a 3:1 or a 3:3 weighting scheme impacts the results when compared to the 3:N case.}

\begin{figure*}
	\includegraphics[width=2.03\columnwidth]{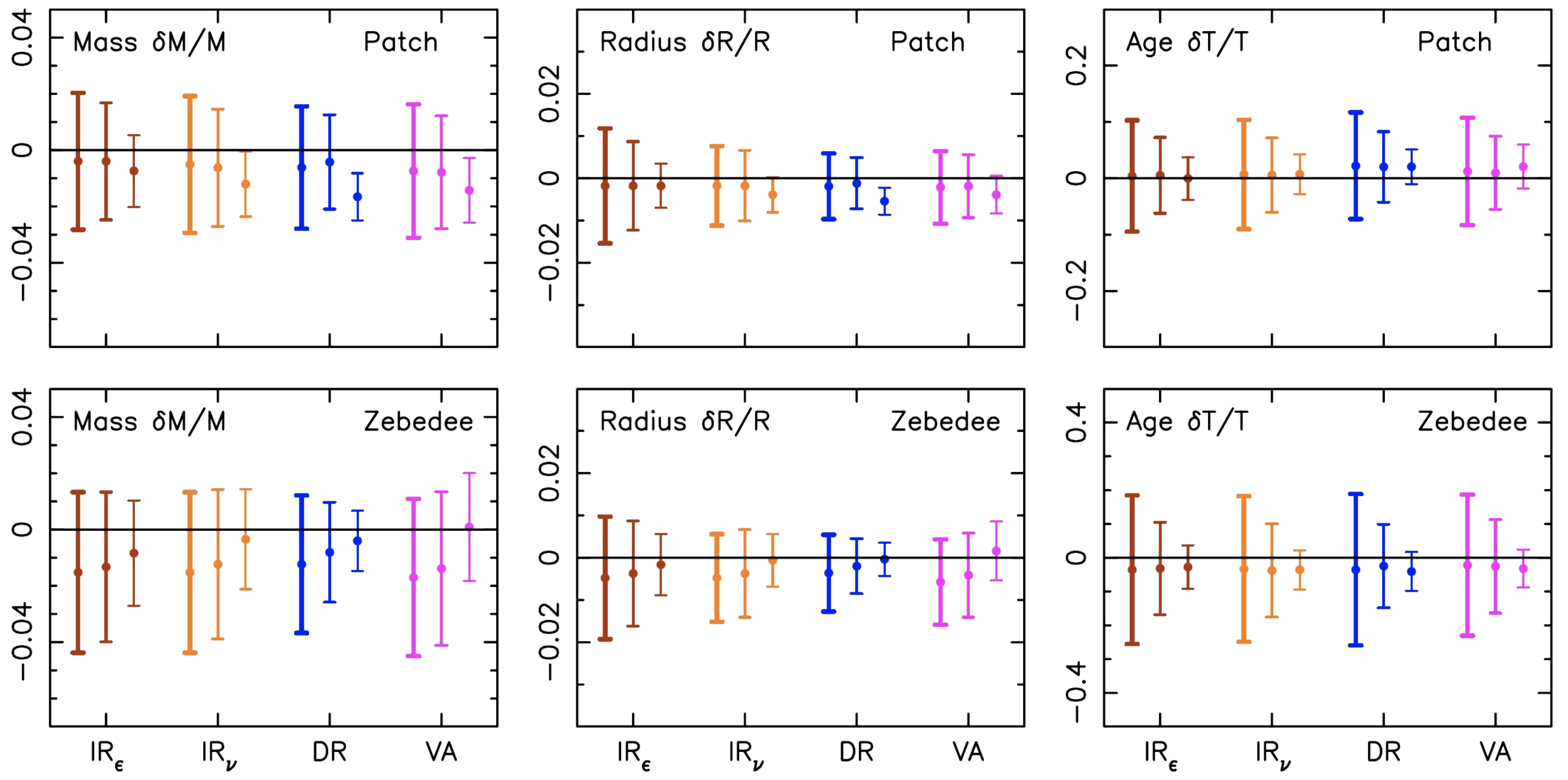}
    \caption{Impact of the weighting scheme on the relative differences between the stellar properties and the corresponding true values. Panels as in Fig.~\ref{fig:comparison}. Results are shown for the two {targets} generated with the same physics as the grid and with parameters within the grid ranges. Four different inference methods are shown, following the colour scheme and Hound ID given in Table~\ref{tab:hounds}. For each inference method, we show results for three different weighting schemes, namely, from thickest to {thinnest} linewidth, 3:1, 3:3, and 3:N.}
   \label{fig:weights}
\end{figure*}

\begin{figure*}
	\includegraphics[width=2.\columnwidth]{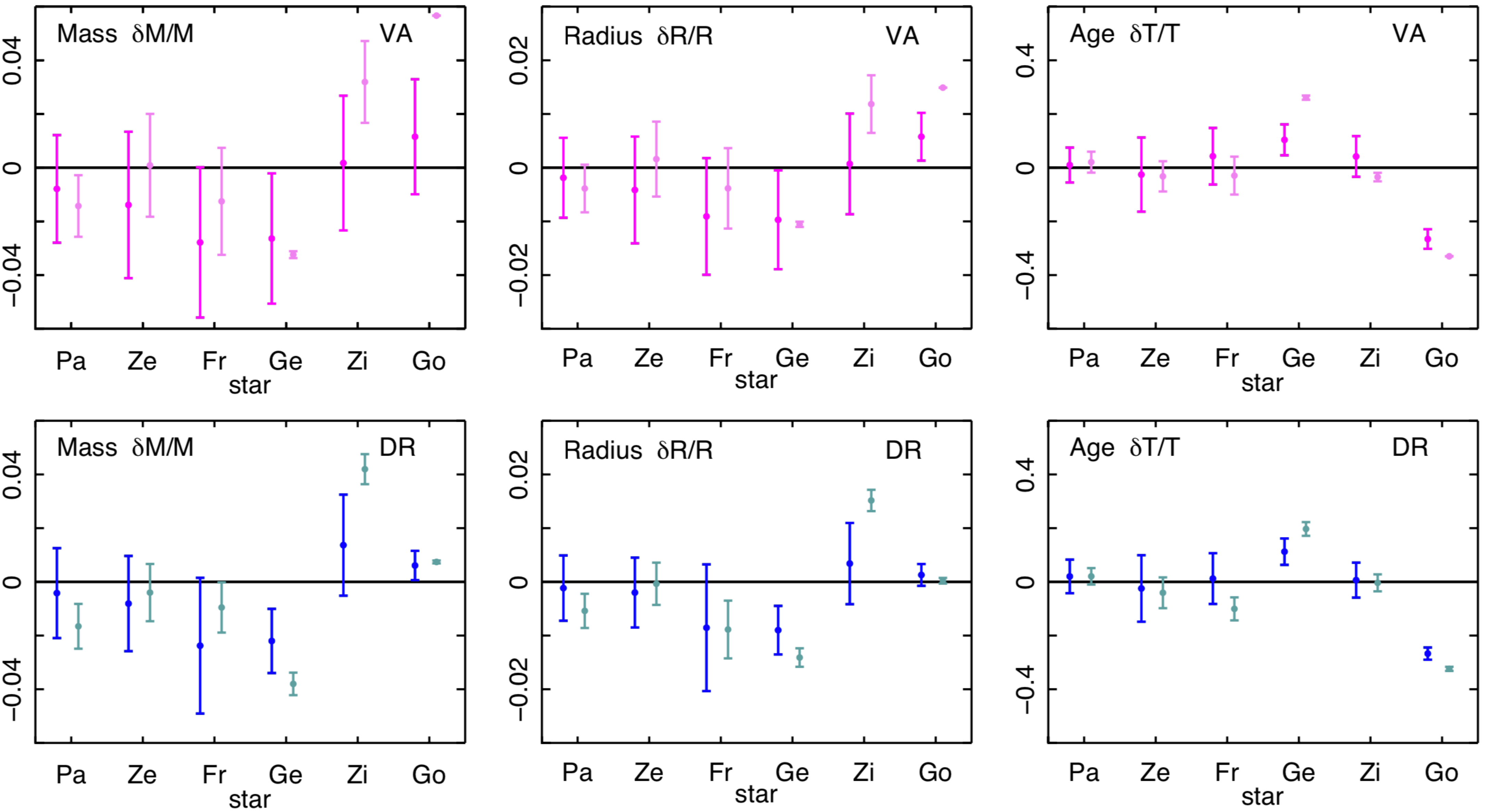}
    \caption{Inferences made by the hounds VA (top panels) and DR (bottom panels) when considering different weights. Panels as in Fig.~\ref{fig:comparison}. For each target and property, two inferences are shown: for a 3:3 weight (left) and a 3:N weight (right). Note the significant decrease in the error bars when each observation is given the same weight ({i.e.} 3:N case), specially for the targets Gerald and Zippy whose physics setup differs from that adopted {while constructing the grid}.}
    \label{fig:weights_dr}
\end{figure*}

{Figures~\ref{fig:weights} and \ref{fig:weights_dr} illustrate how the inferences of the stellar properties are influenced by the weighting scheme. In the first {figure}, the comparison is made for the two stars that fall within the parameter space of the grid and for which the adopted physics is the same as that used in the grid. For these stars, one would expect the true solution to be contained within the grid (even if not corresponding to a grid model) and, thus, the true parameters to be recovered within the statistical errors. The four hounds considered in this exercise were chosen {so} as to cover the most substantial differences in the modelling techniques considered in this study, namely, the use of surface dependent or surface independent methods and different sampling options with {or without  grid interpolation.} }

{Inspection of Fig.~\ref{fig:weights} shows a general decrease in the error bars associated with the inferred properties, as the relative weight of the oscillation frequencies is increased.  This is particularly visible when each observed quantity used in the fits is given the same weight (3:N case), and is a consequence of the problem becoming significantly more constrained when the errors on the frequencies are not inflated.  }

{In addition to the impact on the uncertainties of the inferred properties, the weighting scheme also {slightly} influences  the mean values inferred for each property. In particular, in the case of Patch, the results show that the true values of the mass and radius are outside the $1\sigma$ uncertainties inferred by the hound DR, for the 3:N case. This could result from the statistical errors on the observations (the maximum difference found is only about $2\sigma$). Nevertheless, it is a fact that both the uncertainties and the {inferred} mean values of the properties change differently when changing the weighting, depending on the method applied for the inference. Thus, one may worry that the small statistical error bars inferred in the 3:N case may in some cases be comparable to the differences arising from the different inference procedures or their implementations. }

{The problem becomes more significant if we consider that, unlike in the case for the two targets above, generally the physics adopted to build a grid of models may not fully capture the physics of a real star. In addition to the differences arising from the inference procedures, one would then expect systematic differences resulting from the inadequacy of the models, as discussed, {\it e.g.}, for Gerald in Section~\ref{sec: discrepancies}. {Figure~\ref{fig:weights_dr} illustrates this, by extending the comparison of the 3:3 and 3:N cases to the remaining simulated stars, for the inferences performed by the hounds VA and DR. It is clear that the differences are more significant for the stars whose underlying physics differs from that of the grid, such as Gerald and Zippy. } In these cases, the normalised difference $d_{\rm norm}^{i}$ resulting from equal weighting of the observations (3:N) become significantly larger than 1, with the true values of the stellar properties found many $\sigma$ away from their inferred counterparts. This is likely the reason why the 3:N case is not often used in the context of forward modelling, despite being the only approach built on clear statistical grounds. {Its use in the PLATO pipeline thus requires a complementary and comprehensive study of the systematic errors,  {so} as to ensure a complete characterisation of the uncertainties on the {inferred} properties. Such {a} study is currently ongoing and will be presented in a later work. } }

%\red{[Should perform the case of model averaging and include it in the DR figure for comparison and discussion.]}

\section{{Impact from the length and quality of the data sets}}
\label{sec:length}
%\red{[Ian: Here comes the figure that is missing, comparing the impact of using different data set lengths, from just classical observations to the full data set. Discuss the results based on two hounds (epsilon (IR) and frequencies (IR)) for the same two stars. Highlight importance of having seismic data and of having at least one $l=2$ mode.]}

 {The targets considered so far were produced assuming similar seismic data quality. However, the oscillation mode set returned by the simulations is significantly impacted both by the length of the observations (assuming the quality of the data does not change significantly with time) and the brightness of the target. Reducing the length of the data set, and/or reducing the apparent brightness of a given target will not only reduce the number of modes with returned frequencies but also the precision associated with each frequency. The exact extent of those changes depends on the complex interplay of several factors, including the intrinsic oscillation spectrum, the noise background and frequency resolution, and the observed realization of noise.}
 
 {The study of the impact on the inferred stellar properties from changing the observation length and/or the stars' apparent brightness is beyond the scope of this paper and will be presented in a future work. Here, we address only the impact from changing the set of observed modes and corresponding uncertainties without worrying about the exact underlying cause of those changes. To that end, two exercises were performed, based on the simulations for Patch and Zebedee, the two stars with the same physics as the grid and falling within the grid parameter space. Firstly, we explored the impact on the properties inferred for Patch from decreasing the number of observed modes and the diversity of mode degrees, without modifying the uncertainties in the corresponding mode frequencies. Secondly, we looked at the impact from degrading the quality of the data simulated for Zebedee, with the consequent decrease in the number of observed frequencies and increase in the frequency uncertainties. }
 
\begin{figure*}
	\includegraphics[width=2\columnwidth]{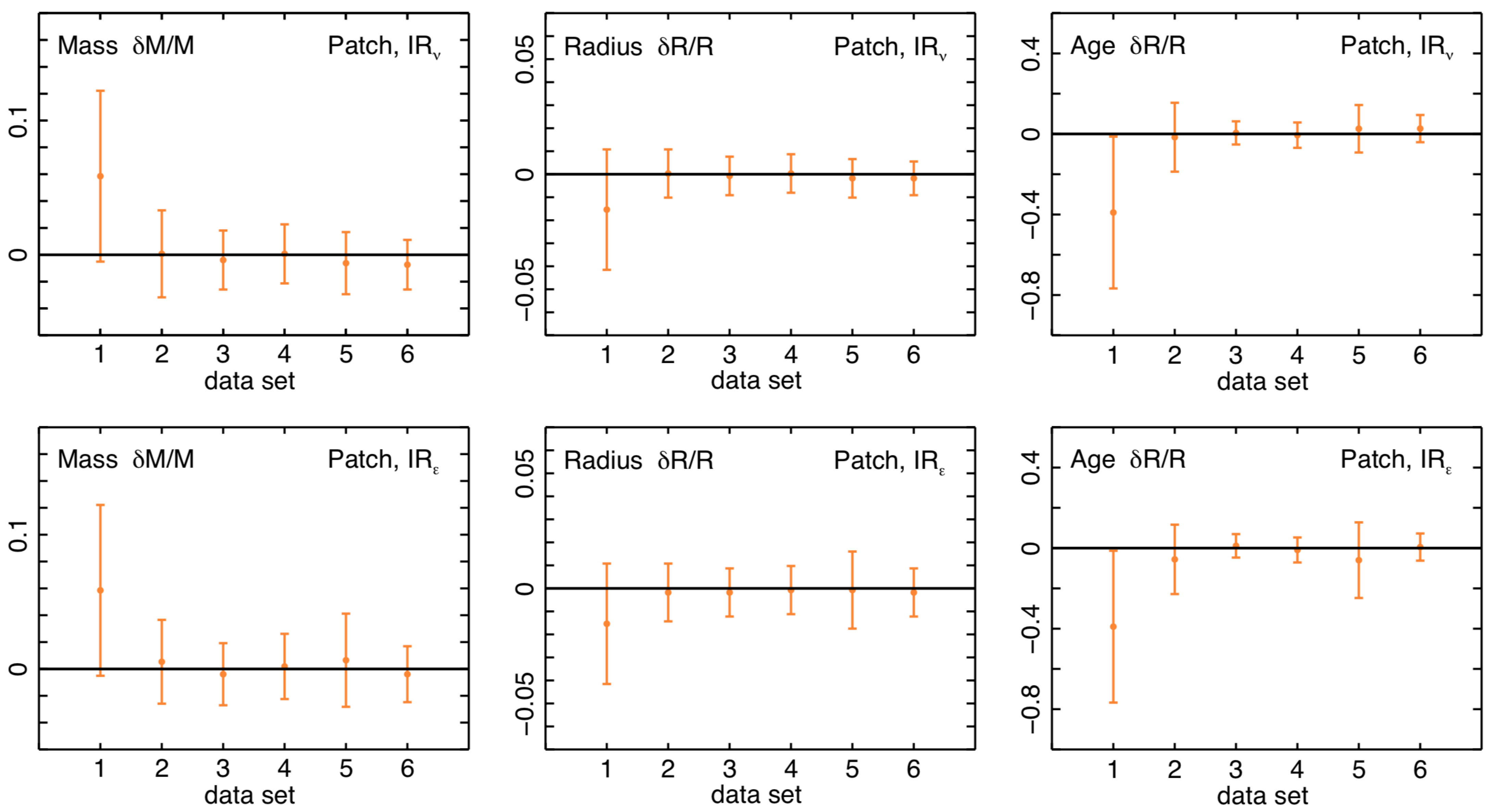}
    \caption{{Inferences made by the hound IR with two methods, IR$_{\nu}$ (top) and IR$_{\epsilon}$ (bottom) for the target Patch,} when fitting different lengths of the seismic data set. Panels as in Fig.~\ref{fig:comparison}. From left to right in each panel, the data sets are composed of $L$, $T_{\rm eff}$, [Fe/H] and the frequency sets listed in Table~\ref{tab:lenght}, namely: 1) No frequencies; 2) 2 $l=0$ and 2 $l=1$ modes; 3) 2 $l=0$ and 2 $l=2$ modes; 4) 2 $l=0$, 2 $l=1$, and 2 $l=2$ modes; 5) 8 $l=0$ and 9 $l=1$ modes; 6) 8 $l=0$, 9 $l=1$, and 6 $l=2$ modes  (full data set). In each data set, the frequencies included were centred on $\nu_{\rm max}$. }
   \label{fig:lenght}
\end{figure*}

 %%% DRR: strange spelling for "lenght" in the figure reference
 {Figure~\ref{fig:lenght} shows the results from the first of these exercises, {performed with the method IR$_\nu$, employing a BG-1term correction, and the method IR$_\epsilon$, both with a 3:3 weight.} The seismic data sets considered in the fits are listed in Table~\ref{tab:lenght}. The improvement resulting from including any set of seismic data is clear for all three stellar properties. Indeed, an increase in both accuracy and precision is seen when comparing the inferences made by fitting data set 1 (no seismic data)  with those made from fits to data sets 2 to 6 (different seismic data combinations). Also striking is the fact that fitting the full set of seismic data or just a small subset of it leads to mass and radius inferences of comparable accuracy and precision.  The situation is somewhat different for the age, where the results show that the inclusion of $l=2$ modes in the data set leads to more precise inferences. This is clearly seen by comparing the uncertainty in the age inferred from fitting data sets 3 (including 2 modes of $l=0$ and 2 modes of $l=2$) with that inferred from fitting data set 5 (including 8 modes of $l=0$ and 9 modes of $l=1$) {and is more evident for the surface-independent method.}  }

\begin{figure*}
	\includegraphics[width=2.0\columnwidth]{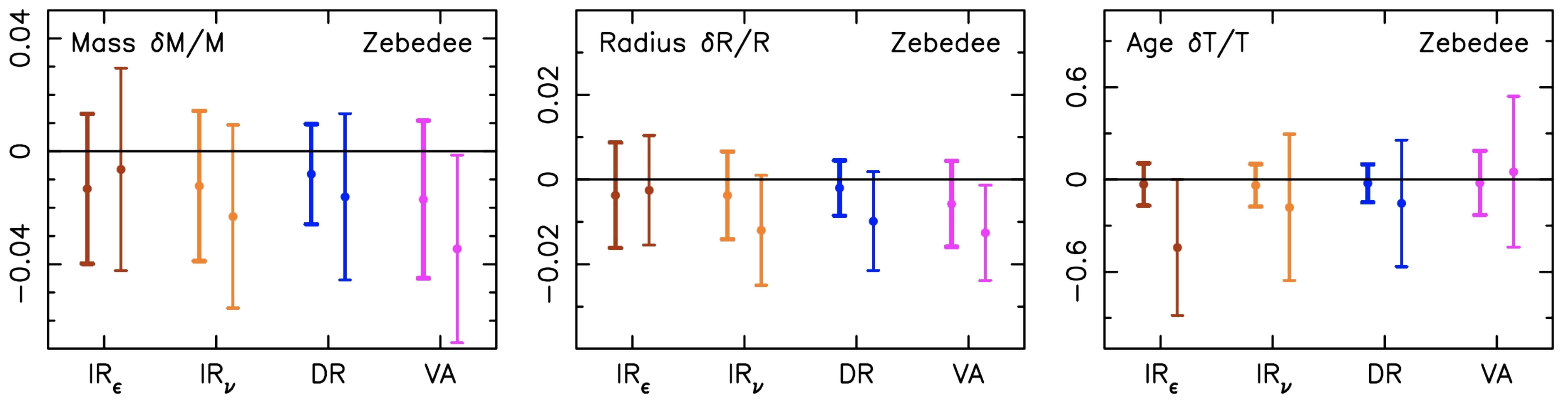}
    \caption{{Comparison of the stellar parameters inferred for Zebedee when considering the original (thicker lines) and degraded (thinner lines) data sets, respectively. Panels as in Fig.~\ref{fig:comparison}.  Four different inference methods are shown, following the colour scheme and Hound ID given in Table~\ref{tab:hounds}. } }
    \label{fig:noise}
\end{figure*}

{The data from the two simulations performed for Zebedee in the context of the second exercise are shown in Tables~\ref{tab:freq_hares} (original data set) and \ref{tab:low_quality} (degraded data set). The number of detected modes decreases from 23 to 7 between the two data sets and the uncertainties in the corresponding frequencies increase by a factor of $\sim$~3.  The impact of these changes is illustrated in Figure~\ref{fig:noise} and Table~\ref{tab:zebedee}. The drastic decrease in the number of modes and associated increase in the uncertainties of the detected frequencies, seems to have a relatively modest impact on the precision and accuracy of the mass and radius inferred for the target. For the precision, a maximum $\sigma_{\rm rel}^i$ of 3.59~per cent and 1.30~per cent is found for the mass and radius, respectively, and for the accuracy, the maximum absolute value of the relative differences, $\max\left(\left|{d_{\rm rel}^i}\right|\right)$, is 3.46~per cent for the mass and of 1.26~per cent for the radius. Nevertheless, the lack of detection of $l=2$ modes in the degraded data set is found to have a significant impact on the seismic constraining power on the age, in accordance with the findings from the first exercise in this section. Moreover, the impact seems to depend on the inference procedure, being significantly greater in the case of the surface independent method (IR$_\epsilon$). While for the hounds based on the fitting of individual frequencies we find a $\max\left(\left|{d_{\rm rel}^i}\right|\right)$ for the age of 18~per cent, the relative difference between the true age and the age inferred from the modelling based on the {surface independent} method is $\left|{d_{\rm rel}^i}\right|\sim 44$~per cent. We note, however, that Zebedee is a relatively young star, having an age of $\sim$ 3 Gyr and a mass comparable to that of the {Sun}. This young age impacts the measure of the age accuracy and precision since their assessment is based on quantities that depend on the inverse of the true age.}

\begin{table}
%	\centering
	\caption{Seismic data fitted in the cases illustrated in Figure~\ref{fig:lenght}. In addition to the seismic data set, the observational constraints included $L$, $T_{\rm eff}$ and [Fe/H], with a 3:3 weight. Columns 2 to 4 show the number of modes of degree $l=0, 1$ and 2 included in each set.  }
	\label{tab:lenght}
	\resizebox{0.8\columnwidth}{!}{%
	\begin{tabular}{lcccl} % four columns, alignment for each
 %     \multicolumn{4}{l}{Zebedee degraded data set}\\
    \hline    
Set & $l=0$ & $l=1$ & $l=2$ & Comments	\\
\hline	
1 & 0  &  0 &  0 & no frequencies\\
2 & 2 & 2  &  0 &\\
3 & 2  & 0  &  2 &\\
4 & 2  & 2  &  2& \\
5 & 8 & 9&   0 &\\
6 & 8  & 9  &  6 & full set\\
\hline		
 	\end{tabular}%
	}
\end{table}

\begin{table*}
	\centering
	\caption{Stellar properties inferred for Zebedee considering {the original (subscript "ori") and degraded (subscript "deg") data sets. The results for the original data set are the same as those given in Tables~\ref{tab:mass} to \ref{tab:age} and are shown here for comparison with the inferences made from the degraded data set}.}
	\label{tab:zebedee}
	\resizebox{2\columnwidth}{!}{%
	\begin{tabular}{llcccccc} % four columns, alignment for each
		\hline
&		&	Mass$_{\rm ori}$	&	Mass$_{\rm deg}$	&	Radius$_{\rm ori}$	&	Radius$_{\rm deg}$	&	Age$_{\rm ori}$	&	Age$_{\rm deg}$	\\
&		&	1.0165	&	1.0165	&	0.9646	&	0.9646	&	3.085	&	3.085	\\
\hline														
{\bf Hounds} &		&		&		&		&		&		&		\\
\hline														
&	Infered value	&	1.003(0.027)	&	1.010(0.037)	&	0.961(0.012)	&	0.962(0.013)	&	2.99(0.42)	&	1.7(1.4)	\\
IR$_\epsilon$ &	$d_{\rm rel}^i$  ($\%$)	&	-1.33	&	-0.64	&	-0.37	&	-0.25	&	-3.18	&	-44.24	\\
&	$d_{\rm norm}^i$ 	&	-0.50	&	-0.18	&	-0.30	&	-0.20	&	-0.23	&	-1.00	\\
&	$\sigma_{\rm rel}^i$  ($\%$)	&	2.66	&	3.59	&	1.24	&	1.29	&	13.71	&	44.14	\\
\hline														
&	Infered value	&	1.004(0.027)	&	0.993(0.033)	&	0.961(0.010)	&	0.953(0.013)	&	2.97(0.43)	&	2.53(1.5)	\\
IR$_ \nu$&	$d_{\rm rel}^i$  ($\%$)	&	-1.23	&	-2.31	&	-0.37	&	-1.20	&	-3.76	&	-18.07	\\
&	$d_{\rm norm}^i$ 	&	-0.46	&	-0.71	&	-0.36	&	-0.92	&	-0.27	&	-0.38	\\
&	$\sigma_{\rm rel}^i$  ($\%$)	&	2.66	&	3.25	&	1.04	&	1.30	&	13.81	&	47.51	\\
\hline														
&	Infered value	&	1.008(0.018)	&	1.000(0.030)	&	0.9627(0.0063)	&	0.955(0.011)	&	3.01(0.38)	&	2.6(1.3)	\\
DR &	$d_{\rm rel}^i$  ($\%$)	&	-0.84	&	-1.61	&	-0.20	&	-0.98	&	-2.46	&	-15.50	\\
&	$d_{\rm norm}^i$ 	&	-0.47	&	-0.55	&	-0.31	&	-0.85	&	-0.20	&	-0.38	\\
&	$\sigma_{\rm rel}^i$  ($\%$)	&	1.77	&	2.95	&	0.65	&	1.16	&	12.38	&	41.16	\\
\hline														
&	Infered value	&	0.999(0.028)	&	0.981(0.034)	&	0.959(0.010)	&	0.952(0.011)	&	3.02(0.64)	&	3.2(1.5)	\\
VA &	$d_{\rm rel}^i$  ($\%$)	&	-1.72	&	-3.46	&	-0.58	&	-1.26	&	-2.20	&	4.99	\\
&	$d_{\rm norm}^i$ 	&	-0.62	&	-1.04	&	-0.57	&	-1.12	&	-0.11	&	0.10	\\
&	$\sigma_{\rm rel}^i$  ($\%$)	&	2.79	&	3.33	&	1.01	&	1.12	&	20.86	&	49.09	\\
\hline														
&	$b_{\rm rel}$ ($\%$)	&	-1.28	&	-2.01	&	-0.38	&	-0.92	&	-2.90	&	-18.20	\\
&	$b_{\rm norm}$ 	&	-0.51	&	-0.62	&	-0.39	&	-0.77	&	-0.20	&	-0.41	\\
&	$\varepsilon_{\rm rel}$ ($\%$)	&	1.32	&	2.25	&	0.40	&	1.01	&	2.96	&	25.24	\\
&	$\varepsilon_{\rm norm}$	&	0.52	&	0.69	&	0.40	&	0.85	&	0.21	&	0.57	\\
&	$\sigma_{\rm rel}$ ($\%$)	&	2.47	&	3.28	&	0.99	&	1.22	&	15.19	&	45.47	\\
\hline
% \multicolumn{10}{l}{*Averages performed excluding the results for George (see text for details).} \\
	\end{tabular}%
	}
\end{table*}

\section{Impact from the errors on the classical parameters}
\label{errors}
%\red{[To solve: Shall we add yet another table for the results in this section? Here the table would show only the relative differences to the true values for each case, along with the measure of the dispersion and absolute difference introduced below.]}

%\red{[To solve: The conclusion that the impact from doubling the uncertainties is negligible was drawn only based on the results from one hound. Should we double-check with another hound?]}

\begin{figure*}
	\includegraphics[width=2\columnwidth]{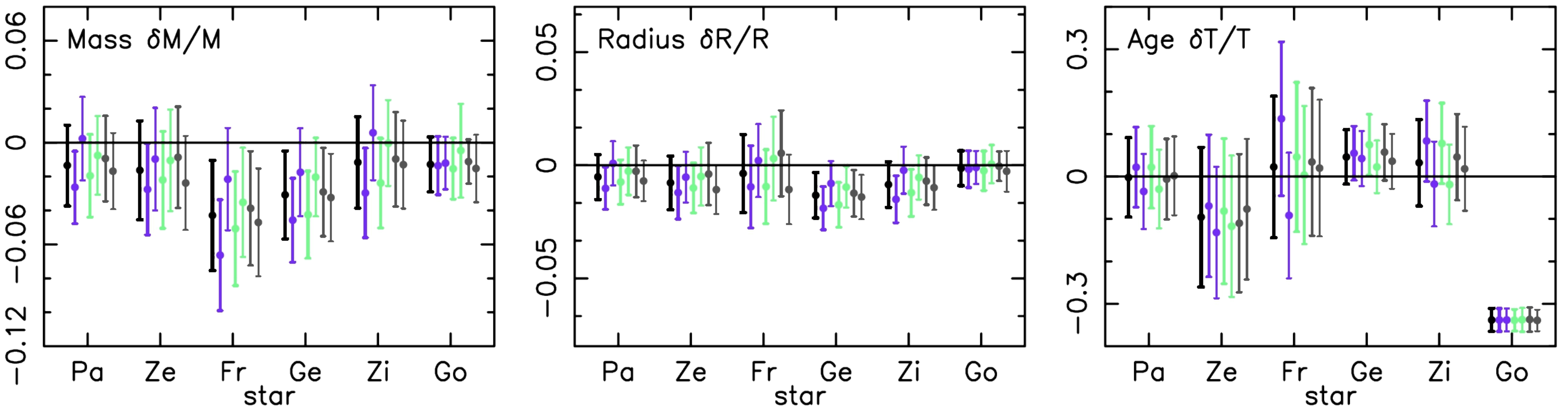}
    \caption{Inferences for a single hound (SB) when a 1$\sigma$ shift in the classical parameters is considered. Panels as in Fig.~\ref{fig:comparison}. For each {target}, seven inferences are shown in the following order, from left to right: 1. results with unchanged classical parameters (black), 2. [Fe/H] shifted by $-1\sigma$ (purple,thick), 3. [Fe/H] shifted by $+1\sigma$  (purple,thin), 4. $L$ shifted by $-1\sigma$ (green, thick), 5. $L$ shifted by $+1\sigma$ (green, thin), 6. $T_{\rm eff}$ shifted by $-1\sigma$ (grey, thick), 7. $T_{\rm eff}$ shifted by $+1\sigma$ (grey, thick).  }
    \label{fig:errors}
\end{figure*}

In this section, we explore the impact on the inferred stellar properties from changing the observational uncertainties on the classical constraints. Two exercises were performed. In the first the 1$\sigma$ uncertainties on [Fe/H], $L$, and $T_{\rm eff}$ were doubled, one at {a} time, while in the second the observed classical constraints were shifted, along with the original errors, by $\mp 1\sigma$.  The impact on the inferred properties from doubling the uncertainties on the classical constraints was found to be generally negligible for the mass and radius, both in terms of the accuracy and the precision of the results, and smaller, for all three properties, than the impact of shifting the classical constraints by $\mp 1\sigma$. The differences found when performing these shifts are illustrated in Fig.~\ref{fig:errors}. Here we show, for the first hound (SB; Table~\ref{tab:hounds}), a comparison between the properties inferred when considering the original classical observations (Table~\ref{tab:classical}; leftmost point in each cluster of results in Fig~\ref{fig:errors}), and classical observations shifted by $\mp 1\sigma$, one at the time (following 6 points in each cluster of results). 

Inspection of Fig.~\ref{fig:errors} shows that the impact on the accuracy and precision of the inferred parameters from shifting the classical observations by $\mp 1\sigma$ is generally small, but not negligible, particularly in the case of the age.  Using as a reference the relative difference {with} the true value obtained with the classical observations given in Table~\ref{tab:classical}, ($d_{\rm rel}^{\rm ref}$, black in Fig~\ref{fig:errors}), we computed the dispersion of the results for each property and target as before: $\sqrt{\sum_i(d_{\rm rel}^i-d_{\rm rel}^{\rm ref})^2/N_{\rm c}}$, where $N_{\rm c}=6$ is the number of cases considered. Moreover, we also computed the maximum departure between any given inference and the value inferred in the reference case, $\left | d_{\rm rel}^i-d_{\rm rel}^{\rm ref} \right |$. The maximum dispersion is found for Fred, with values of 1.39~per cent, 0.68~per cent, and 6.7~per cent for mass, radius, and age, respectively. The maximum departure from the reference result is also found for Fred. In the case of the mass and age, this maximum is found for shifts in [Fe/H], with values of 2.34~per cent and 11.5~per cent, respectively. For the radius, we find a maximum departure of 0.89~per cent, arising from a shift in $T_{\rm eff}$.

\section{Conclusions}
\label{sec:conclusions}
%\red{[Summary of the main results. Comparison of results with those in other studies in the literature, including the SPACEINN H$\&$H but also studies on the impact of changing the surface correction prescription, the errors on the classical constraints, etc. ]}

In this work we compared different approaches to the asteroseismic inference of stellar properties based on a pre-computed grid of models and corresponding pulsation properties. The aim was to understand the accuracy and precision that may be expected on the inferred properties when applying state-of-the-art techniques and identify critical aspects of the inference process that may require further development, in light of the preparation for the soon to be launched PLATO mission, from ESA. 
The study was conducted based on a single grid of models and six main-sequence artificial stars, three of which {were} generated with the same physics setup as the grid (although one of these has an enrichment ratio outside the range covered by the grid parameter space). The remaining three stars were generated with at least one aspect of the {underlying} physics differing from the physics adopted for the grid. {Five different grid-based inference methods, namely SB, JO, DR, IR$_\epsilon$ and VA, and two variants of these, have been compared. The methods are summarised in appendix~\ref{apB} and Table~\ref{tab:hounds}}.

With regards to the comparison between different grid-based inference methods, our main conclusions can be summarised as follows:
\begin{itemize}
    \item No significant differences were found among the methods elected for comparison with regards to the accuracy of the results, when these are considered in light of the reference values of 15, 2 and 10 per cent in mass, radius and age, respectively. Specifically, considering the 5 targets for which the hounds reported results they could trust ({i.e.,} excluding George), the average relative errors on the inferences made with the 5 different methods varied in the interval 1.61~--~2.61 per cent for the mass, 0.33~--~0.84 per cent for the radius and 5.11~--~6.01 per cent for the age. 
    
    \item Similarly, the differences in precision on the mass among the elected methods was not deemed significant. Considering the same 5 stars, the average precision on mass varied within the interval 1.52~--~2.75 per cent among the 5 methods,  in all cases being much smaller than 15 per cent. For the radius, the average precision of the methods was found to vary within the interval 0.58~--~1.23 per cent and for the age within the interval 4.96~--~11.82 per cent. While these differences may seem more significant, we note that to a large extent they result from the relative weight on the classical and seismic constraints adopted by each hound for the elected method. As expected, methods applying a 3:1 weight generally have larger error bars than those applying a 3:3 weight {(where 3:1/3:3 indicates that the full set of frequencies 
is given the same weight as one/three global constraints, respectively; cf. Sections~\ref{sec:differeces} and \ref{weight})}. However,  this weight is not intrinsic to the method (in the sense that different weights can be adopted with the same method). Thus, the differences seen in the precision of the {inferred} properties do not translate {into} a fundamental difference in the potential precision of the methods themselves. In addition, the significant average age precision of 4.96 per cent found with the method by the hound JO resulted in part from neglecting the perturbation of the frequencies in the MC simulations, as noted in Section~\ref{sec:pdfs}. When considering the same weight (Fig.~\ref{fig:weights}) the precision of different methods on age was found to be similar. For the radius, the  method employed by DR seems to be the most precise and that by IR$_\epsilon$ the least precise, with the error bars on the latter found to be up to a factor of $\sim$2 larger than those on the former.
\end{itemize}

Concerning the impact of the ad hoc choices that may be involved in the inference procedures, such as those associated with the surface corrections and the relative weight set on the classical and seismic constraints, we reached the following conclusions:
\begin{itemize}
\item If surface corrections are not added to the model frequencies when these are used directly in the fits, the relative differences between the inferred and true values of the stellar properties are very significant. For the method by DR, on which the surface corrections tests were based, the relative differences were found to be as large as $\sim$ 7, 3 and 35 per cent for mass, radius and age, respectively, in the absence of a surface correction. The inclusion of a BG-2term surface correction in this method reduces the maximum of the relative differences on mass, radius and age to 2.36, 0.9 and 11.25 per cent. Still, the choice of the prescription for the surface corrections was found to impact the results, leading to a maximum dispersion on the inferences of 1.9, 1.0 and 6.8 per cent in mass, radius and age, respectively. While these values of the dispersion are smaller than the reference values, they are by no means negligible in the case of the radius and age. Thus, this result calls for an improvement of the modelling of the surface layers of stars, both in what concerns the structure and the pulsations \citep{mosumgaard20,belcacem21,jorgensen21}.

\item Given a set of observations, the adoption of a weight in the fitting procedure aimed at decreasing the relative impact of the seismic data with regards to the classical data is equivalent to an ad hoc inflation of the errors on the frequencies. To be statistically sound, the inference method to be used in the PLATO pipeline should instead give each observation the same weight (our 3:N case). However, our results show that when the targets do not share the physics setup of the grid, as will generally happen for real stars, the properties inferred with a 3:N approach can be many sigma away from the true property values. This is mostly related {to} the fact that a 3:N approach leads to significantly smaller uncertainties on the inferred properties when the grid does not contain a reasonable sample of comparably good models around the inferred solution.  This result points to an urgent need to thoroughly characterise the systematic errors incurred on the inferred stellar properties when performing inferences based on a grid similar to {the one} to be adopted by the PLATO mission. These systematic errors, resulting from fixing a given set of options concerning the physics of the grid, need to be considered along with the formal errors derived from the application of the inference procedure, in order to provide robust uncertainties on the inferred properties of PLATO stars. In some cases, our results also  show a non-negligible change in accuracy when comparing the 3:3 and 3:N weights, both for methods with and without interpolation between grid models. Further studies should be pursued to understand these differences, and in particular, to investigate whether they are connected to the grid resolution.
\end{itemize} 

Finally, we have tested the impact of degrading the classical and seismic data. 
%In the case of the classical data we have considered the impact of doubling the uncertainties and that of shifting by 1-$\sigma$ the point estimate values.  In the case of the seismic data, we tested the impact of decreasing the number of individual frequencies and of limiting the diversity of mode degrees, as well as the impact of simultaneously increasing the uncertainties and decreasing the number of detected modes. 
With regards to these tests our conclusions were as follows:

\begin{itemize} 
\item Concerning the classical data, the most significant impact was found when shifting the central values.  Specifically, when changing $L$,$T_{\rm eff}$ and [Fe/H] by $\pm$1\,$\sigma$, one at {a} time, the dispersion in the inferred relative differences  reached up to  1.39, 0.68 and 6.7 per cent in  mass, radius and age, respectively. Moreover, the maximum difference between any two mass or age inferences was found when shifting the value of [Fe/H] and reached 2.34 per cent for mass and 11.5 per cent for age. For the radius the maximum difference was found when shifting $T_{\rm eff}$ and did not exceed 1 per cent. These results highlight the importance of determining the classical parameters to a high precision and accuracy, particularly when considering the impact they { have when inferring} the stellar age.

\item Concerning the seismic data, our results show that the detection of only a small number of oscillation frequencies may be enough to set stringent constraints on the stellar mass and radius. While that seems to be true also for the age, in this case we found that the precision of the inferences depends more strongly on the combination of mode degrees available for the fit, with the results becoming more precise when at least one $l=2$ mode is detected. When in addition to reducing the number of modes and eliminating the modes of degree $l=2$, the uncertainties in the mode frequencies are increased, the inference of a precise and accurate age starts to be compromised. It is, therefore, important to investigate thoroughly the case of stars in the regime where seismic data becomes limited and the inference approach eventually changes from fitting individual frequencies or $\epsilon_l$ phases to fitting global seismic constraints.

\end{itemize} 

It is worth noting that the conclusions summarised here are based on the study of targets whose physics is relatively standard. However, even in the case of low mass stars, some non-standard processes may have a significant evolutionary impact.  An example are macroscopic and microscopic processes leading to chemical transport in radiative regions inside stars \citep[see][for a review]{aerts21}, that, together, dictate the observed surface abundances at a given time in evolution. As illustrated by our study of Gerald, considerable biases in mass and age can result from not accounting for atomic diffusion. In stars slightly more massive than Gerald (and in particular for F stars), the contribution of radiative accelerations to atomic diffusion becomes non-negligible \citep{deal18} and even in relatively slow rotators, rotationally-induced mixing may become an important effect counteracting atomic diffusion \citep{deal20}. These effects, neglected in standard models, may lead to additional biases in the inferred stellar properties, not considered in the present work. This emphasises the need to continue developing a new generation of stellar evolution codes and to acquire data on pulsating stars that may help constrain further these aspects of the physics.

The results presented in this work provide guidance for the development of the PLATO pipeline where it concerns the inference of the properties of stars with seismic data and the characterisation of the associated exoplanetary systems. Moreover, the work identifies additional paths of research that should be pursued in order to achieve the PLATO goals and optimise the science return of the mission.

\section*{Acknowledgements}
This work presents results from the European Space Agency (ESA) space mission
PLATO. The PLATO payload, the PLATO Ground Segment and PLATO data processing
are joint developments of ESA and the PLATO Mission Consortium (PMC). Funding for
the PMC is provided at national levels, in particular by countries participating in the
PLATO Multilateral Agreement (Austria, Belgium, Czech Republic, Denmark, France,
Germany, Italy, Netherlands, Portugal, Spain, Sweden, Switzerland, Norway, and United
Kingdom) and institutions from Brazil. Members of the PLATO Consortium can be found
at \url{https://platomission.com/}. The ESA PLATO mission website is
\url{https://www.cosmos.esa.int/plato}. We thank the teams working for PLATO for all their work.
MSC acknowledges the support by FCT/MCTES through the research grants UIDB/04434/2020, UIDP/04434/2020 and PTDC/FIS-AST/30389/2017, and by FEDER 
%- Fundo Europeu de Desenvolvimento Regional 
through COMPETE2020 
%- Programa Operacional Competitividade e Internacionalização 
(grant: POCI-01-0145-FEDER-030389). MSC and TC are supported by national funds through FCT in the form of work contracts (CEECIND/02619/2017 and CEECIND/00476/2018, respectively). JLR acknowledges support from the Carlsberg Foundation (grant agreement CF19-0649). Funding for the Stellar Astrophysics Centre is provided by The Danish National Research Foundation (Grant DNRF106). {This article made use of AIMS, a software for fitting stellar pulsation data, developed in the context of the SPACEINN network, funded by the European Commission’s Seventh Framework Programme.} BN acknowledges funding from the Alexander von Humboldt Foundation and "Branco Weiss fellowship -- Science in Society" through the SEISMIC stellar interior physics group. DRR, M-JG, KB, RMO acknowledge the support of the French space agency (CNES). A.S. acknowledge support from 
the European Research Council Consolidator Grant funding scheme (project ASTEROCHRONOMETRY, G.A. n. 772293).
%, http://www.asterochronometry.eu).
%\red{[Please add your acknowledgements here]}

%%%%%%%%%%%%%%%%%%%%%%%%%%%%%%%%%%%%%%%%%%%%%%%%%%
\section*{Data Availability}
%%%%%%%%%%%%%%%%%%%%%%%%%%%%%%%%%%%%%%%%%%%%%%%%%%
 
The simulated data used here are available in Tables~\ref{tab:classical}, \ref{tab:freq_hares}, and \ref{tab:freq_hares_2}.


\bibliographystyle{mnras}
\bibliography{solar-like} % if your bibtex file is called example.bib

%\begin{thebibliography}{99}
%\end{thebibliography}

%%%%%%%%%%%%%%%%%%%%%%%%%%%%%%%%%%%%%%%%%%%%%%%%%%

%%%%%%%%%%%%%%%%% APPENDICES %%%%%%%%%%%%%%%%%%%%%

\appendix

\section{Data sets produced by the Hares}
\label{apA}

{Tables~\ref{tab:freq_hares}-\ref{tab:freq_hares_2} list the properties of the individual modes simulated for the targets in Table~\ref{tab:hares}, specifically: {the mode degree, $l$, the radial order, $n$, the true mode frequency value, $\nu^{\rm true}$, the simulated frequency, $\nu$, and its uncertainty, $\sigma_\nu$.} Table~\ref{tab:low_quality} lists the set of degraded observations simulated for Zebedee for the exercise in Section~\ref{sec:length}.}

\begin{table*}
%	\centering
	\caption{{Simulated frequencies for the targets in Table~\ref{tab:hares}. Columns show the mode degree and radial order, followed by the true model frequency, the simulated frequency, and the associated 1\,$\sigma$ error.}}
	\label{tab:freq_hares}
	\resizebox{0.63\columnwidth}{!}{%
	\begin{tabular}{lcccc} % five columns, alignment for each
\multicolumn{5}{l}{Patch}\\									
\hline									
$l$	&	$n$	&	$\nu^{\rm true}$	&	$\nu$	&	$\sigma_\nu$	\\
	&		&	($\mu$Hz)	&	($\mu$Hz)	&	($\mu$Hz)	\\
\hline									
0	&	16	&	2343.51	&	2343.23	&	0.31	\\
0	&	17	&	2475.44	&	2475.35	&	0.20	\\
0	&	18	&	2608.14	&	2608.24	&	0.16	\\
0	&	19	&	2741.55	&	2741.40	&	0.16	\\
0	&	20	&	2874.71	&	2874.55	&	0.18	\\
0	&	21	&	3007.94	&	3008.00	&	0.23	\\
0	&	22	&	3141.48	&	3141.43	&	0.38	\\
0	&	23	&	3275.42	&	3274.02	&	0.73	\\
1	&	15	&	2272.92	&	2273.13	&	0.36	\\
1	&	16	&	2405.01	&	2404.83	&	0.21	\\
1	&	17	&	2537.98	&	2537.87	&	0.15	\\
1	&	18	&	2671.46	&	2671.57	&	0.13	\\
1	&	19	&	2805.19	&	2805.21	&	0.14	\\
1	&	20	&	2938.88	&	2939.08	&	0.17	\\
1	&	21	&	3072.45	&	3071.81	&	0.25	\\
1	&	22	&	3206.60	&	3206.53	&	0.43	\\
1	&	23	&	3341.21	&	3340.91	&	0.93	\\
2	&	16	&	2468.27	&	2468.04	&	0.33	\\
2	&	17	&	2601.55	&	2601.61	&	0.26	\\
2	&	18	&	2735.51	&	2735.41	&	0.25	\\
2	&	19	&	2869.20	&	2869.31	&	0.28	\\
2	&	20	&	3002.96	&	3002.43	&	0.37	\\
2	&	21	&	3136.95	&	3136.34	&	0.59	\\
-	&	-	&	-	&	-	&	-	\\
-	&	-	&	-	&	-	&	-	\\
-	&	-	&	-	&	-	&	-	\\
-	&	-	&	-	&	-	&	-	\\
-	&	-	&	-	&	-	&	-	\\
-	&	-	&	-	&	-	&	-	\\
-	&	-	&	-	&	-	&	-	\\
-	&	-	&	-	&	-	&	-	\\
-	&	-	&	-	&	-	&	-	\\
\hline									
 	\end{tabular}%
	}
		\resizebox{0.63\columnwidth}{!}{%
	\begin{tabular}{lcccc} % five columns, alignment for each
\multicolumn{5}{l}{Zebedee}\\									
\hline									
$l$	&	$n$	&	$\nu^{\rm true}$	&	$\nu$	&	$\sigma_\nu$	\\
	&		&	($\mu$Hz)	&	($\mu$Hz)	&	($\mu$Hz)	\\
\hline									
0	&	18	&	2795.03	&	2794.98	&	0.29	\\
0	&	19	&	2936.71	&	2936.89	&	0.19	\\
0	&	20	&	3079.03	&	3078.92	&	0.16	\\
0	&	21	&	3221.73	&	3221.60	&	0.16	\\
0	&	22	&	3364.06	&	3364.43	&	0.20	\\
0	&	23	&	3506.45	&	3506.28	&	0.30	\\
0	&	24	&	3649.01	&	3647.87	&	0.54	\\
0	&	25	&	3791.98	&	3789.75	&	1.14	\\
1	&	17	&	2720.28	&	2720.07	&	0.34	\\
1	&	18	&	2861.65	&	2861.36	&	0.20	\\
1	&	19	&	3003.99	&	3004.19	&	0.14	\\
1	&	20	&	3146.59	&	3146.30	&	0.13	\\
1	&	21	&	3289.29	&	3289.14	&	0.15	\\
1	&	22	&	3431.82	&	3431.59	&	0.20	\\
1	&	23	&	3574.24	&	3574.49	&	0.33	\\
1	&	24	&	3717.20	&	3718.28	&	0.64	\\
1	&	25	&	3860.49	&	3857.98	&	1.48	\\
2	&	18	&	2924.95	&	2924.71	&	0.31	\\
2	&	19	&	3067.60	&	3067.29	&	0.26	\\
2	&	20	&	3210.61	&	3210.91	&	0.26	\\
2	&	21	&	3353.27	&	3353.16	&	0.32	\\
2	&	22	&	3495.98	&	3496.49	&	0.47	\\
2	&	23	&	3638.78	&	3637.81	&	0.82	\\
-	&	-	&	-	&	-	&	-	\\
-	&	-	&	-	&	-	&	-	\\
-	&	-	&	-	&	-	&	-	\\
-	&	-	&	-	&	-	&	-	\\
-	&	-	&	-	&	-	&	-	\\
-	&	-	&	-	&	-	&	-	\\
-	&	-	&	-	&	-	&	-	\\
-	&	-	&	-	&	-	&	-	\\
-	&	-	&	-	&	-	&	-	\\
\hline									
 	\end{tabular}%
	}
			\resizebox{0.63\columnwidth}{!}{%
		\begin{tabular}{lcccc} % five columns, alignment for each
\multicolumn{5}{l}{Fred}\\									
\hline									
$l$	&	$n$	&	$\nu^{\rm true}$	&	$\nu$	&	$\sigma_\nu$	\\
	&		&	($\mu$Hz)	&	($\mu$Hz)	&	($\mu$Hz)	\\
\hline									
0	&	13	&	971.45	&	971.05	&	0.94	\\
0	&	14	&	1037.84	&	1036.91	&	0.77	\\
0	&	15	&	1104.53	&	1105.40	&	0.66	\\
0	&	16	&	1172.63	&	1171.94	&	0.59	\\
0	&	17	&	1242.10	&	1242.45	&	0.54	\\
0	&	18	&	1312.39	&	1312.31	&	0.52	\\
0	&	19	&	1382.37	&	1382.99	&	0.53	\\
0	&	20	&	1451.27	&	1451.34	&	0.56	\\
0	&	21	&	1519.49	&	1518.40	&	0.63	\\
0	&	22	&	1587.67	&	1587.10	&	0.76	\\
0	&	23	&	1656.26	&	1656.52	&	1.00	\\
0	&	24	&	1725.79	&	1728.67	&	1.47	\\
1	&	12	&	934.62	&	935.19	&	0.91	\\
1	&	13	&	1001.68	&	1001.87	&	0.74	\\
1	&	14	&	1067.86	&	1067.13	&	0.62	\\
1	&	15	&	1135.10	&	1133.87	&	0.53	\\
1	&	16	&	1203.81	&	1202.85	&	0.48	\\
1	&	17	&	1273.76	&	1274.52	&	0.45	\\
1	&	18	&	1344.08	&	1344.35	&	0.44	\\
1	&	19	&	1413.64	&	1413.34	&	0.46	\\
1	&	20	&	1482.22	&	1482.53	&	0.50	\\
1	&	21	&	1550.46	&	1550.98	&	0.58	\\
1	&	22	&	1618.89	&	1618.46	&	0.72	\\
1	&	23	&	1688.00	&	1687.72	&	1.00	\\
2	&	14	&	1099.01	&	1097.51	&	1.07	\\
2	&	15	&	1166.93	&	1167.08	&	0.95	\\
2	&	16	&	1236.26	&	1235.29	&	0.87	\\
2	&	17	&	1306.51	&	1307.35	&	0.84	\\
2	&	18	&	1376.56	&	1375.87	&	0.84	\\
2	&	19	&	1445.59	&	1445.21	&	0.89	\\
2	&	20	&	1513.89	&	1514.92	&	0.99	\\
2	&	21	&	1582.14	&	1582.95	&	1.19	\\
\hline																
 	\end{tabular}%
	}
\end{table*}

\begin{table*}
%	\centering
	\caption{Simulated frequencies for the targets in Table~\ref{tab:hares} (cont.).}
	\label{tab:freq_hares_2}
	\resizebox{0.63\columnwidth}{!}{%
	\begin{tabular}{lcccc} % five columns, alignment for each
\multicolumn{4}{l}{Gerald}\\									
\hline									
$l$	&	$n$	&	$\nu^{\rm true}$	&	$\nu$	&	$\sigma_\nu$	\\
	&		&	($\mu$Hz)	&	($\mu$Hz)	&	($\mu$Hz)	\\
\hline									
0	&	15	&	1706.25	&	1706.47	&	0.24	\\
0	&	16	&	1807.84	&	1807.47	&	0.15	\\
0	&	17	&	1910.09	&	1910.07	&	0.11	\\
0	&	18	&	2013.33	&	2013.11	&	0.10	\\
0	&	19	&	2116.36	&	2116.48	&	0.11	\\
0	&	20	&	2219.24	&	2219.29	&	0.12	\\
0	&	21	&	2322.51	&	2322.75	&	0.16	\\
0	&	22	&	2425.98	&	2425.89	&	0.24	\\
0	&	23	&	2529.90	&	2529.59	&	0.46	\\
1	&	14	&	1649.67	&	1650.16	&	0.29	\\
1	&	15	&	1751.59	&	1751.46	&	0.16	\\
1	&	16	&	1854.03	&	1854.06	&	0.11	\\
1	&	17	&	1957.00	&	1957.02	&	0.09	\\
1	&	18	&	2060.39	&	2060.37	&	0.09	\\
1	&	19	&	2163.93	&	2164.05	&	0.09	\\
1	&	20	&	2267.15	&	2267.16	&	0.11	\\
1	&	21	&	2370.70	&	2370.64	&	0.16	\\
1	&	22	&	2474.93	&	2474.73	&	0.27	\\
1	&	23	&	2579.29	&	2579.56	&	0.58	\\
2	&	15	&	1801.62	&	1801.50	&	0.24	\\
2	&	16	&	1904.23	&	1904.16	&	0.19	\\
2	&	17	&	2007.81	&	2008.10	&	0.17	\\
2	&	18	&	2111.20	&	2111.20	&	0.17	\\
2	&	19	&	2214.50	&	2214.53	&	0.19	\\
2	&	20	&	2318.14	&	2318.34	&	0.25	\\
2	&	21	&	2421.96	&	2421.70	&	0.38	\\
2	&	22	&	2526.28	&	2527.13	&	0.72	\\
-	&	-	&	-	&	-	&	-	\\
-	&	-	&	-	&	-	&	-	\\
-	&	-	&	-	&	-	&	-	\\
-	&	-	&	-	&	-	&	-	\\
-	&	-	&	-	&	-	&	-	\\
-	&	-	&	-	&	-	&	-	\\
-	&	-	&	-	&	-	&	-	\\
-	&	-	&	-	&	-	&	-	\\
\hline									
 	\end{tabular}%
	}
		\resizebox{0.63\columnwidth}{!}{%
	\begin{tabular}{lcccc} % five columns, alignment for each
\multicolumn{4}{l}{Zippy}\\									
\hline									
$l$	&	$n$	&	$\nu^{\rm true}$	&	$\nu$	&	$\sigma_\nu$	\\
	&		&	($\mu$Hz)	&	($\mu$Hz)	&	($\mu$Hz)	\\
\hline									
0	&	13	&	1238.22	&	1239.08	&	0.53	\\
0	&	14	&	1324.78	&	1324.53	&	0.37	\\
0	&	15	&	1410.83	&	1411.18	&	0.28	\\
0	&	16	&	1495.27	&	1494.80	&	0.23	\\
0	&	17	&	1579.81	&	1580.39	&	0.21	\\
0	&	18	&	1665.40	&	1665.50	&	0.20	\\
0	&	19	&	1752.11	&	1752.00	&	0.21	\\
0	&	20	&	1838.94	&	1838.99	&	0.24	\\
0	&	21	&	1925.46	&	1925.26	&	0.30	\\
0	&	22	&	2011.72	&	2012.07	&	0.45	\\
0	&	23	&	2098.20	&	2098.64	&	0.78	\\
1	&	12	&	1189.99	&	1189.87	&	0.59	\\
1	&	13	&	1276.40	&	1276.66	&	0.38	\\
1	&	14	&	1363.08	&	1362.97	&	0.27	\\
1	&	15	&	1448.41	&	1448.50	&	0.22	\\
1	&	16	&	1532.90	&	1533.33	&	0.19	\\
1	&	17	&	1617.90	&	1617.86	&	0.17	\\
1	&	18	&	1704.40	&	1704.86	&	0.17	\\
1	&	19	&	1791.40	&	1791.68	&	0.19	\\
1	&	20	&	1878.46	&	1878.61	&	0.22	\\
1	&	21	&	1965.00	&	1964.84	&	0.30	\\
1	&	22	&	2051.67	&	2051.53	&	0.48	\\
1	&	23	&	2138.62	&	2139.27	&	0.91	\\
2	&	13	&	1318.06	&	1318.56	&	0.60	\\
2	&	14	&	1404.34	&	1404.56	&	0.45	\\
2	&	15	&	1488.97	&	1489.17	&	0.37	\\
2	&	16	&	1573.63	&	1574.06	&	0.33	\\
2	&	17	&	1659.29	&	1659.30	&	0.32	\\
2	&	18	&	1746.20	&	1746.07	&	0.33	\\
2	&	19	&	1833.26	&	1832.56	&	0.38	\\
2	&	20	&	1920.09	&	1920.62	&	0.48	\\
2	&	21	&	2006.63	&	2007.57	&	0.70	\\
2	&	22	&	2093.40	&	2093.48	&	1.21	\\
-	&	-	&	-	&	-	&	-	\\
-	&	-	&	-	&	-	&	-	\\
\hline											
 	\end{tabular}%
	}
			\resizebox{0.63\columnwidth}{!}{%
		\begin{tabular}{lcccc} % five columns, alignment for each
\multicolumn{4}{l}{George}\\									
\hline									
$l$	&	$n$	&	$\nu^{\rm true}$	&	$\nu$	&	$\sigma_\nu$	\\
	&		&	($\mu$Hz)	&	($\mu$Hz)	&	($\mu$Hz)	\\
\hline									
0	&	13	&	986.72	&	986.52	&	0.44	\\
0	&	14	&	1055.70	&	1055.59	&	0.28	\\
0	&	15	&	1125.65	&	1125.73	&	0.21	\\
0	&	16	&	1194.98	&	1194.94	&	0.17	\\
0	&	17	&	1263.30	&	1263.60	&	0.15	\\
0	&	18	&	1331.12	&	1330.98	&	0.14	\\
0	&	19	&	1399.92	&	1399.97	&	0.15	\\
0	&	20	&	1469.69	&	1469.46	&	0.16	\\
0	&	21	&	1540.26	&	1540.18	&	0.20	\\
0	&	22	&	1610.67	&	1610.93	&	0.28	\\
0	&	23	&	1680.94	&	1680.96	&	0.44	\\
0	&	24	&	1750.73	&	1750.91	&	0.85	\\
1	&	12	&	951.51	&	951.58	&	0.50	\\
1	&	13	&	1019.70	&	1019.58	&	0.30	\\
1	&	14	&	1089.90	&	1089.83	&	0.20	\\
1	&	15	&	1159.98	&	1159.97	&	0.16	\\
1	&	16	&	1229.28	&	1229.26	&	0.14	\\
1	&	17	&	1297.42	&	1297.56	&	0.13	\\
1	&	18	&	1366.01	&	1365.98	&	0.12	\\
1	&	19	&	1435.54	&	1435.38	&	0.13	\\
1	&	20	&	1506.05	&	1505.87	&	0.15	\\
1	&	21	&	1576.72	&	1576.79	&	0.20	\\
1	&	22	&	1647.11	&	1647.16	&	0.29	\\
1	&	23	&	1717.08	&	1717.01	&	0.51	\\
1	&	24	&	1786.73	&	1787.51	&	1.09	\\
2	&	13	&	1051.31	&	1051.04	&	0.47	\\
2	&	14	&	1121.39	&	1120.90	&	0.34	\\
2	&	15	&	1190.88	&	1190.94	&	0.27	\\
2	&	16	&	1259.38	&	1259.06	&	0.24	\\
2	&	17	&	1327.27	&	1327.30	&	0.23	\\
2	&	18	&	1396.12	&	1395.78	&	0.24	\\
2	&	19	&	1465.91	&	1466.19	&	0.26	\\
2	&	20	&	1536.54	&	1536.09	&	0.32	\\
2	&	21	&	1607.01	&	1606.78	&	0.43	\\
2	&	22	&	1677.32	&	1676.46	&	0.69	\\
\hline									
 	\end{tabular}%
	}
\end{table*}

\begin{table}
%	\centering
	\caption{Zebedee degraded simulated data set.}
	\label{tab:low_quality}
	\resizebox{0.49\columnwidth}{!}{%
	\begin{tabular}{lccc} % four columns, alignment for each
 %     \multicolumn{4}{l}{Zebedee degraded data set}\\
    \hline    
        \multicolumn{4}{l}{$L/L_\odot$ = 1.00 $\pm$ 0.03}\\
        \multicolumn{4}{l}{$T_{\rm eff}$ = 5887 $\pm$ 85 K}\\
        \multicolumn{4}{l}{[Fe/H] =  -0.03 $\pm$ 0.09 dex}\\
        \multicolumn{4}{l}{$\nu_{\rm max}$ = 3260 $\pm$ 167 $\mu$Hz}\\
        \multicolumn{4}{l}{$\Delta\nu$ = 143.1 $\pm$ 2.8 $\mu$Hz}\\
    \hline
$l$	&	$n$	&	$\nu$	&	$\sigma_\nu$	\\
	&		&	($\mu$Hz)	&	($\mu$Hz)	\\
\hline	
0 & 21  &  3223.09  &  0.48\\
0 & 22  & 3364.38  &  0.63\\
1 & 19  & 3004.18  &  0.44\\
1 & 20  & 3147.48  &  0.39\\
1 & 21 &  3289.17 &   0.45\\
1 & 22  & 3433.58  &  0.65\\
1 & 23  & 3576.01  &  1.17\\
\hline		
 	\end{tabular}%
	}
\end{table}

\section{Methods employed by the Hounds}
\label{apB}

\subsection{SB: The grid method}

This method used the provided grid as-is, without any interpolation. The first step in the analysis of each {target} was to select the subset of models that lay within $\pm 8\sigma$ of {$\Delta\nu$}, $\nu_{\rm max}$, $T_{\rm eff}$, luminosity and [Fe/H] of each star. While a $\pm 6\sigma$-cut is generally sufficient, the $8\sigma$ cut allowed us to use the same set of models for the tests where the spectroscopic parameters were shifted by 1$\sigma$.

The frequencies of each of the selected models were corrected for surface effects using the \citet{ball14} two-term correction. The parameters for the correction were determined using radial modes and then applied to all modes. The corrected frequencies were then used to calculate the $\chi^2$ per degree of freedom, which we call $\chi^2_{\nu}$, for each model. This is defined as:
\begin{equation}
    \chi^2(\nu)=\frac{1}{N-1}{\sum\left({\frac{\nu_{nl}^{\rm obs}-\nu_{nl}^{\rm corr}}{\sigma^{\rm obs}_{nl}}}\right)^2},
\label{eq:eqchinu}
\end{equation}
where $\sigma^{\rm obs}_{nl}$ is the uncertainty on the frequency ${\nu_{nl}}$ of the {target}, and {the sum is over all $N$ observed frequencies}. This is then used to calculate a likelihood
\begin{equation}
    {\mathcal L}(\nu)=C\exp\left(-\frac{\chi^2(\nu)}{2}\right),
    \label{eq:nulike}
\end{equation}
$C$ being the normalisation constant.

{The surface term correction does not take into account the fact that the frequency difference between the model and the star is expected to be {smaller} at low frequencies than at high frequencies. There are many models in the grid that are different enough from the star that the high-frequency modes match but the low-frequency modes do not. Additionally,  contrary to the expectations, the low-frequency modes of these models have lower frequencies than that of the star. To ensure that these models are given a lower weight than others, we also calculated the $\chi^2$ value for the 2 lowest uncorrected model frequencies for all available degrees. We divide this by 10000 to reduce this term's contribution to the likelihood, call it $\chi^2_{\rm low}$, and calculate a weight that is  defined as}
\begin{equation}
{\mathcal W}={W}\exp\left(-\chi^2_{\rm {low}}\right),
\end{equation}
{where W is a normalisation constant.}
%\DRR{(Shouldn't there be a normalization constant to ensure that the sum of $\mathcal{W}$ over all models is 1?)}
Since ${\mathcal W}$ is normalised such that its sum over all models is 1, the division by 10000 results in a gentle down selection.

As with the corrected frequencies, we calculate likelihoods for $T_{\rm eff}$, luminosity and [Fe/H]. For example, the likelihood for effective temperature was defined as
\begin{equation}
{\mathcal L}(T_{\rm eff})=D\exp(-\chi^2(T{_{\rm eff}})/2),
\label{eq:tcal}
\end{equation}
with
\begin{equation}
\chi^2(T_{\rm eff})=\frac{(T^{\rm obs}_{\rm eff}-T^{\rm model}_{\rm eff})^2}{\sigma^2_{ T}},
\label{eq:chit}
\end{equation}
where $\sigma_{T}$ is the uncertainty on the effective temperature, and $D$ the normalisation constant. We define the likelihoods for [Fe/H] and $L$ in a similar manner.

The total likelihood for each model is then
\begin{equation}
{\mathcal L}_{\rm total}={\mathcal W}{\mathcal L}(\nu){\mathcal L}(T_{\rm eff}){\mathcal L}({[\rm{Fe/H]}}){\mathcal L}(L).
\end{equation}
The  {means} of the marginalised likelihoods of the ensemble of models {were} used to determine the parameters of the
star, after converting them to a probability density by normalising the likelihood by the prior distribution of the property.

\subsection{JO: Grid Monte Carlo}
\label{Jo}

We use the same Monte-Carlo grid search procedure as in \cite{ong21}, but with a different set of penalty functions. For each model, an overall cost function $\chi^2_\text{tot}$ is computed as the sum of the following contributions:
\begin{itemize}
    \item $\chi^2_\text{global} = \sum_i \left(y_{i,\text{model}} - y_{i,\text{obs}} \over \sigma_i\right)^2$, where $y$ are global quantities: we have used the classical spectroscopic constraints, as well as  $\nu_\text{max}$ and $\epsilon_c$ \citep[the radial mode phase offset at $\nu_\text{max}$, in the sense of][]{ong19}.
    \item $\chi^2_\text{BG}$, being the reduced-$\chi^2$ penalty function from applying the surface correction of \cite{ball14}. The parameters are fitted against only the radial modes, and then the cost function is computed from applying the correction to all mode frequencies.
    \item $\chi^2_\epsilon$, which is the reduced $\chi^2$ function of the $\epsilon_l$-matching algorithm described in \cite{roxburgh16} --- cf. their eq. 12 and discussion in §\ref{IRe}.
    \item $\chi^2_\text{low n}$: under the ansatz that the surface term affects higher-order modes more than it does low-order ones, we construct a quantity ${1 \over N}\sum_n^N \left[J\left(\nu_{n,0,\text{model}} - \nu_{n,0,\text{obs}} \over \sigma_{n,0}\right)\right]^2$, summing over only the $N$  lowest-frequency radial modes. $J$ is an asymmetric penalty function satisfying
    \begin{equation}
        J(x) = \left\{\begin{array}{cc}
            x & x < 0 \\
            x/f & x \ge 0
        \end{array}\right.,
    \end{equation}
    for some constant $f$. This is to penalise surface corrections which modify low-frequency modes too much, as well as to ensure that the sense of the surface term is the same as seen in the case of the solar surface term. For this exercise we have used $N = 5, f = 10$. We moreover downweighted this term by a factor of $1/4$, to ensure that the only function it serves is that of regularisation.
\end{itemize}

For each model $m$, we treat the quantity $w_m = \exp \left[-\chi^2_\text{tot}/2\right] / p_m$, normalised to sum to 1 over all models in the grid, as a likelihood weight function. Here $p_m$ is the grid's sampling density function, which may not be uniform, and is not assumed to be known in advance. We estimate $p_m$ using a kernel density estimate applied to the grid parameters (including model ages). We then compute the weighted mean of each of the output quantities with respect to these likelihood weights $w_m$, summing over all models in the grid. This corresponds to taking the posterior means of the desired output properties, under the assumption of uniform priors on the model parameters.

We repeat the same procedure, perturbing each of the spectroscopic and average seismic inputs by a normally-distributed random amount with variance given by the observational errors. Unless otherwise indicated, we do not perturb the frequencies, since evaluating the cost terms associated with individual mode frequencies is extremely expensive. We collect these posterior means associated with each realisation, over $10^4$ realisations. The posterior means define a mapping $g: y \mapsto \theta$, where $y$ are the supplied observations, and $\theta$ the desired output properties (e.g., mass, radius, etc). By perturbing the input observations in this manner, we directly propagate "input" probability distributions on $y$ to "output" probability distributions on $\theta$ under the action of the map $g$. We report the marginalised medians and quantiles as our estimates for the values and uncertainties of the output parameters with respect to these output probability distributions.

\subsection{DR: AIMS}

The Asteroseismic Inference on a Massive Scale (AIMS) code takes in a grid of models and applies a Markov Chain Monte Carlo (MCMC) approach to fitting a given set of classic and seismic constraints. In order to allow the MCMC approach to explore the parameter space more freely, AIMS carries out interpolation within the input grid of models {using a} triangulation (or tessellation) {of the parameter space} between the evolutionary tracks, and linear interpolation along the tracks.  This gives it a great degree of flexibility in terms of the location of the evolutionary tracks in parameter space.  The interpolation is applied both to model properties such as mass, radius, and age, and to the pulsation frequencies.  This then provides all of the necessary information to calculate the priors and likelihood function which intervene in the probability calculations.  Once the MCMC run is completed, a representative sample of models is obtained from which it is possible to calculate posterior probability distributions for the models properties, as well as statistical averages, standard deviations, correlations, and various percentiles.  For more details on the AIMS code, we refer the reader to \citet{rendle19} as well as to the AIMS documentation\footnote{The latest version of AIMS is available at: https://gitlab.com/sasp/aims}.

In the present hare-and-hounds exercise, AIMS was used to fit the hares using a variety of different settings.   In terms of surface corrections, the approaches by \citet{kjeldsen08}, \citet{ball14}, and \citet{sonoi15} were used.  In each case, variants with one and two free parameters were used (in particular, the second parameter for the \citealt{kjeldsen08} surface correction is the $b$ exponent and that of the \citealt{sonoi15} correction is the $\beta$ exponent, both of which are fitted non-linearly with the MCMC approach).  {Runs without surface corrections were also carried.}
%as were runs using the frequency ratios $r_{01} + r_{10}$, $r_{02}$, and their combination, supplemented with the lowest frequency radial mode as well as the average large frequency separation based on radial modes.
A $3\sigma$ cutoff on classic constraints was applied, except when applying the two-term \citet{ball14} surface correction where various cutoffs where applied: $1\sigma$, $3\sigma$, $5\sigma$, $\infty$. {With the exception of George in the specific case of a $1\sigma$ cutoff, the differences in the results for different cutoffs were found to be negligible. Thus, only the $3\sigma$ cutoff results are shown.} {In all cases we used uniform priors on the relevant ranges of the parameters. In the case of the age, a uniform prior over the interval [0, 13.8]~Gyr was considered.}
Various weights on classic and seismic constraints were adopted for the runs, namely 3:3, 3:N, and in a few cases 3:1.  The constraint on $\nu_{\max}$ was not included but it is possible to include it.
% I wonder why I forgot the constraint on nu_max.  Perhaps Benard inlucded it in his runs.

\subsection{IR$_\epsilon$: surface independent}
\label{IRe}

%\DRR{(In this whole section, it would be good to use a consistent notation for the indices $n,\ell$, ie. do you include a comma, a decimal point, or nothing between the two.)}
Epsilon matching is a ``surface layer independent'' model comparison algorithm which subtracts the contribution of the outer layers of the stars from a combination of their frequencies.
%This algorithm and others  such as comparing separation ratios, rest on the   (very accurate) approximation that the contribution of the 
%outer layers depends only on frequency and not on angular degree.
The epsilon matching algorithm used here is described in detail in \cite{roxburgh16}. 

In short, the (adiabatic) oscillation frequencies of a star can be expressed in terms of epsilons, $\epsilon_{nl}$, as
$$\nu_{nl} = \Delta\left(  n+ {l\over 2} + \epsilon_{nl} \right)~~{\rm so}~~~~
\epsilon_{nl} =\epsilon_l(\nu_{nl}) = {\nu_{nl}\over\Delta}   - n - {l\over 2}
%= (\nu_{n,\ell}\over\Delta} -n - {\ell\over 2}
$$
where $\Delta$ is an estimate of the large separation. 
Moreover, the $\epsilon _l(\nu)$ can be expressed in terms of an $l$-dependent inner phase shift determined by the inner structure of the star, $\delta_l(\nu)$, and an $l$-independent outer phase shift, $\alpha(\nu)$, determined by the structure of the outer layers, such that  $\epsilon _l(\nu) =\alpha(\nu) +\delta_l(\nu)$.
%where $\alpha(\nu)$ is the $\ell$ independent outer phase shift and  $\delta_\ell(\nu)$ is the $\ell$ dependent inner phase shift.
 
The algorithm then consists of the following steps:
\begin{itemize}
\item Determine the epsilons of the observed star  $\epsilon_l^{o}(\nu_{nl}^{o})$, and associated errors $ s_{nl}={\sigma_{nl}^o/\Delta}$;
 
\item Determine the epsilons of the model $\epsilon^{m}(\nu_{nl}^{m})$;

\item Interpolate in $\epsilon^{m}(\nu^{m})$ for $\epsilon^{m}(\nu^{o})$ and determine  ${\cal E} (l, \nu^o_{nl})=
\epsilon^m_l(\nu^o_{nl})-\epsilon^o_l(\nu^o_{nl})$ and form
%$$\chi^2_\epsilon = \sum_{n,\ell} \left({\cal E} (\ell, \nu^o_{n\ell}) - {\cal F}(\nu^o_{n\ell})\over s_{n\ell} \right)^2. \eqno(12) \label{eq:chi2_ir}$$
%$$ {1\over N-M} \sum_1^N \left( {\cal E}(\ell, \nu^o_{n\ell}) - {\cal F}(\nu^o_{n\ell})\over s_{n\ell} \right)^2 \eqno(12)$$
%
\begin{equation}
    \hspace{2cm}    \chi^2_\epsilon = \sum_{nl} \left({\cal E} (l, \nu^o_{nl}) - {\cal F}(\nu^o_{nl})\over s_{nl} \right)^2.
   \label{eq:chi2_ir}
\end{equation}
%\end{center}
The function ${\cal F}(\nu)$ subtracts the $l$-independent contribution from the outer layers and is an 
$M$ parameter $l$-independent  function of $\nu$.
%with the parameters determined by minimising $\chi^2_\epsilon$ per degree of freedom with respect to $M$, with $M$ constrained to be less than the number of $l=0$ frequencies. 
{The functional form of ${\cal F}(\nu)$ used here is
\begin{equation}
{\cal F}(\nu)=\sum_{k=0}^MC_k T_k\left(\zeta\right),
\end{equation}
where $\zeta=(\nu^o-\nu^o_{\rm min})/(\nu^o_{\rm max}-\nu^o_{\rm min})$ and the $T_k$ are Chebychev polynomials of order $k$. The coefficients $C_k$ and the upper limit $M$ are determined by the condition that the $\chi^2$ per degree of freedom is minimised, with $M$ constrained to be less than the number of $l=0$ frequencies. The value of $M$ is not necessarily the same for fits to different models.}

\item Determine $\chi^2_s=\sum_i \left(y_{i,\text{model}} - y_{i,\text{obs}} \over \sigma_i\right)^2$, where $y$ are the classical parameters, here $L$, $T_{\rm eff}$ and [Fe/H], and $\sigma_i$ the corresponding uncertainties;

\item Also, if desired,  determine $\chi^2_0$,  defined as above, but with a single term where $y$ is the frequency of the lowest $l=0$ mode.~~ 
%with some  weight  here determined by the average values of $\chi^2_0$ and $\chi^2_e$ over model set.

\item Add $\chi^2_s, \chi^2_\epsilon$, $\chi^2_0$ applying different weights, as desired, to obtain the likelihood function, then the probability density distributions,  mean, standard deviation and percentiles on mass, radius and age.
\end{itemize}

\subsection{VA: BASTA}
%\red{[Victor: please clarify which priors were used on the properties.]}

The BAyesian STellar Algorithm \citep[{\tt BASTA},][]{2015MNRAS.452.2127S,BASTA}\footnote{The code is available at https://github.com/BASTAcode/BASTA} is a fitting pipeline written in Python designed to determine stellar properties combining photometric, spectroscopic, astrometric, and asteroseismic observations. The code uses a grid of stellar models and Bayesian statistics to compute the marginalised posterior distribution of any desired quantity present in the grid by comparing its predicted values {with}  a pre-defined combination of input observed properties.

%For the present exercise we ran {\tt BASTA} in the following configurations: one set of results was determined fitting individual frequencies and adopting the correction of \citet{ball14} to account for the surface effect, and a second set of stellar properties was obtained by fitting the frequency ratios $r_{012}$ as recommended by \citet{2018arXiv180807556R}. In both cases we also included the effective temperature, photospheric luminosity, and surface abundance ratio [Fe/H] in the fitted quantities. We adopted weights of 3:N, 3:3, and 3:1 between the atmospheric and asteroseismic constraints to test their impact.

For the present exercise we ran {\tt BASTA} in the following configuration: {we fitted individual frequencies and adopted the two-term correction of \citet{ball14} to account for the surface effect. We also included the effective temperature, photospheric luminosity, and surface abundance ratio [Fe/H] in the fitted quantities.} We adopted weights of 3:N, 3:3, and 3:1 between the atmospheric and asteroseismic constraints to test their impact.

{In the case of the method VA$_{\rm int}$,} to increase the resolution of the original grid we performed interpolations across and along evolutionary tracks as described in \cite{BASTA}. Briefly, for each target we selected tracks within a broad range encompassing the observed large frequency separation, effective temperature, and metallicity. The resolution is then increased across tracks by a multiplicative factor in all parameters used to construct the grid (mass, initial [Fe/H], initial helium abundance, mixing-length, and overshooting efficiency). The new tracks are found via a tessellation of these base parameters using {\tt scipy.spatial.Delaunay}. For this exercise we adopted a multiplicative factor of 20,
%\red{[MC: the files I received say level10 which, from your README file, seems to me as indicating a factor of 10. I thus wrote 10xGrid in Table 3. Should that be corrected to 20xGrid, or is the value of 20 the one that is incorrect?]}
resulting in an increase of the number of tracks in the range of interest for each target from $\sim$700 to $\sim$15,000. The new tracks are then interpolated along the tracks to increase the resolution in frequency using {\tt scipy.interpolate.interp1d}. Between two consecutive models in the track we required a variation smaller than $1\mu$Hz in the lowest observed $l=0$ mode.

\subsection{IR$_\nu$: surface dependent}
This method compares observed and model frequencies and classical parameters with a Ball and Gizon  correction \citep{ball14} added to the model frequencies. For a given observed frequency set one reads in the properties, frequencies and mode inertias of the models
and adds a Ball and Gizon correction to the model frequencies, where the  coefficients are determined so as to minimise $\chi^2_\nu$ of the fit of corrected model to observed frequencies and 
not in terms of the fundamental properties of the model/star, where
$$\chi^2_\nu = \sum \left( \nu_{nl}^{\rm modcorr}  -\nu^{\rm obs}_{nl}  \over \sigma^{\rm obs}_{nl} \right)^2.
$$
It is straightforward to take different prescriptions for the ``corrections'',  {as well as no corrections}.

Likewise, one determines $\chi^2_s=\chi^2_L+\chi^2_T+\chi^2_F$ defined as in \ref{IRe}. The search is in general limited to the volume with $\chi^2_L, \chi^2_T, \chi^2_F < 9$ but in some cases  to $< 25$.

One determines a total $\chi^2 =w_s \chi^2_s + w_\nu  \chi^2_\nu$
where $w_s$ and $w_\nu$ are prescribed weights ({\it e.g.} the 3:3 results shown in Sec.~\ref{sec:comparison} correspond to $w_s=1, w_\nu=3/N$ where $N$ is the number of frequencies). Taking $w_s=1, w_\nu=1$ corresponds to giving equal weight to each individual frequency and each classical parameter. 

Given the values of the $\chi^2$ this in turn gives the likelihoods and hence the probability density functions and the means, standard deviations, and percentiles of the model parameters over the set of models included in the analysis.

%%%%%%%%%%%%%%%%%%%%%%%%%%%%%%%%%%%%%%%%%%%%%%%%%%

% Don't change these lines
\bsp	% typesetting comment
\label{lastpage}
\end{document}